\documentclass[aps,pre,floats,floatfix,twocolumn,superscriptaddress]{revtex4-1}

\usepackage{latexsym,amsmath}
\usepackage{amsbsy}
\usepackage[pdftex]{graphicx}
\usepackage{amssymb}
\usepackage{epstopdf}
\usepackage[usenames]{color}
\usepackage{float}
\usepackage{graphicx}
\usepackage[normalem]{ulem}

\begin{document}

\title{Multiscale unfolding of real networks by geometric renormalization}

\author{Guillermo Garc\'{\i}a-P\'erez}
\affiliation{Departament de F{\'\i}sica de la Mat\`eria Condensada, Universitat de Barcelona, Mart\'{\i} i Franqu\`es 1, 08028 Barcelona, Spain}
\affiliation{Universitat de Barcelona Institute of Complex Systems (UBICS), Universitat de Barcelona, Barcelona, Spain}

\author{Mari\'an Bogu\~n\'a}
\affiliation{Departament de F{\'\i}sica de la Mat\`eria Condensada, Universitat de Barcelona, Mart\'{\i} i Franqu\`es 1, 08028 Barcelona, Spain}
\affiliation{Universitat de Barcelona Institute of Complex Systems (UBICS), Universitat de Barcelona, Barcelona, Spain}

\author{M. \'Angeles Serrano}
\affiliation{Departament de F{\'\i}sica de la Mat\`eria Condensada, Universitat de Barcelona, Mart\'{\i} i Franqu\`es 1, 08028 Barcelona, Spain}
\affiliation{Universitat de Barcelona Institute of Complex Systems (UBICS), Universitat de Barcelona, Barcelona, Spain}
\affiliation{ICREA, Pg. Llu\'is Companys 23, E-08010 Barcelona, Spain}

\date{\today}

\begin{abstract}
Multiple scales coexist in complex networks. However, the small world property makes them strongly entangled. This turns the elucidation of length scales and symmetries a defiant challenge. Here, we define a geometric renormalization group for complex networks and use the technique to investigate networks as viewed at different scales. We find that real networks embedded in a hidden metric space show geometric scaling, in agreement with the renormalizability of the underlying geometric model. This allows us to unfold real scale-free networks in a self-similar multilayer shell which unveils the coexisting scales and their interplay. The multiscale unfolding offers a basis for a new approach to explore critical phenomena and universality in complex networks, and affords us immediate practical applications, like high-fidelity smaller-scale replicas of large networks and a multiscale navigation protocol in hyperbolic space which boosts the success of single-layer versions.
\end{abstract}

\maketitle

\section{Introduction}
Symmetries permeate reality and our theories to understand it. From very simple to very subtle, all of them denote invariance under a transformation, and thus similarity or even exact correspondence between different parts of a system or between the system and itself when observed at different scales of length, or other variable. As paradigmatic examples, fractals are geometric objects showing physical scale invariance and self-similarity~\cite{Mandelbrot:1961yi}. Moreover, these properties can also apply to phenomenological behaviours like systems dynamics near critical points of phase transitions~\cite{Stanley:1971zf}. 

In complex networks, multiple scales coexist but they are so entangled that the definition of self-similarity and scale-invariance has been limited by the lack of a valid source of geometric length scale transformations. Previous efforts to study these symmetries are based on topology and include coarse-graining to preserve the large-scale behaviour of random walks~\cite{PhysRevLett.99.038701}, or box-covering procedures based on shortest path lengths between nodes~\cite{Song:2005uq,Goh:2006,Song:2006,Kim:2007,PhysRevLett.101.148701,PhysRevLett.104.025701}. The latter revealed that certain real networks have finite fractal dimensions and exhibit self-similarity, although scaling in the topological properties was not observed beyond the degree distribution and the maximum and average degrees. However, the collection of shortest paths, albeit a well-defined metric, is a poor source of length-based scaling factors in networks due to the small-world~\cite{Watts:1998ga} or even ultrasmall-world~\cite{Cohen:2003rh} property, and the problem remained controversial. Other studies have faced the multiscale structure of network models in a somewhat more geometric way~\cite{Newman1999341,Boettcher2011}, but their findings cannot be directly applied to real-world networks.

The development in the last years of plausible models of complex networks based on an underlying metric space~\cite{Serrano:2008hb,BoPa10} opens now the door to a proper geometric definition of self-similarity and scale invariance and to an unfolding of the different scales present in the connectivity structure of real networks. Hidden metric space network models couple the topology of a network to an underlying geometry through a universal connectivity law which combines popularity and similarity dimensions~\cite{Serrano:2008hb,KrPa10,Papadopoulos:2012uq}, such that more popular and similar nodes have more chance to interact. Naturally, the geometricalization of networks allows a reservoir of distance scales so that we can borrow concepts and techniques from the renormalization group in statistical physics~\cite{Kadanoff,Wilson:1975}, which has been used to study systems where widely different length scales are present simultaneously. By recursive averaging over short-distance degrees of freedom, the renormalization group has successfully explained, for instance, the universality properties of critical behavior in phase transitions~\cite{Wilson:1983}.

In this work, we introduce a geometric renormalization group for complex networks (RGN). The method is based on a geometric embedding of the networks to construct renormalized versions of their structure by coase-graining neighbouring nodes into supernodes and defining a new map which progressively selects longer range connections by identifying relevant interactions at each scale. The RGN technique is inspired by the block spin renormalization group devised by L. P. Kadanoff ~\cite{Kadanoff}.

\section{Evidence of geometric scaling in real networks}
The map of a complex network embedded in a hidden metric space, $\mathcal{M}(T,G)$, contains information about both its topology $T$ and geometry $G$ (in terms of the positions of the nodes in the hidden metric space). Given $\mathcal{M}(T,G)$, we define a geometric renormalization operator $\mathbb{F}_r$ of resolution $r$ which coarse-grains the original network by a factor $r$ and defines a new topology $T'$ and a new geometry $G'$ conforming the renormalized map $\mathcal{M}'$
\begin{equation}
\mathcal{M}(T,G) \stackrel{\mathbb{F}_r}{\longrightarrow} \mathcal{M}'(T',G').
\end{equation}
The transformation zooms out by changing the minimum length scale from that of the original network to a larger value. This operation can be iterated starting from the original network at $l=0$,
\begin{equation}
\mathcal{M}^{(l+1)}(T^{(l+1)},G^{(l+1)}) =\mathbb{F}_r [\mathcal{M}^{(l)}(T^{(l)},G^{(l)})]. 
\end{equation}
In the limit $N \rightarrow \infty$, it can be applied up to any desired scale of observation, whereas it is bounded to $\mathcal{O}(\log N)$ iterations in systems with a finite number of nodes $N$.

\begin{figure}[h!]
\centering
\includegraphics[width=0.9\linewidth]{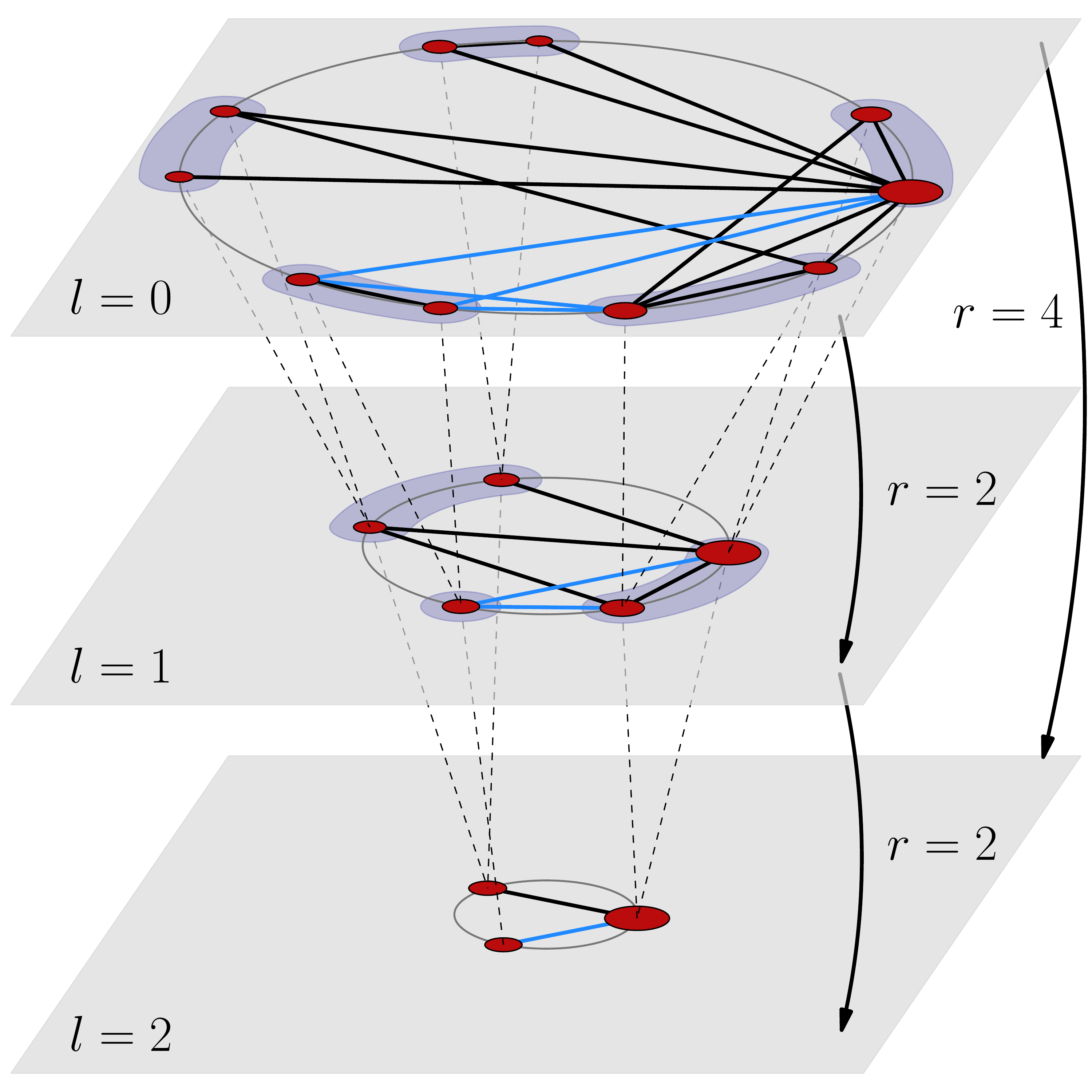}
\caption{{\bf Geometric renormalization transformation for complex networks.} Each layer is obtained after a renormalization step with resolution $r$ starting from the original network in $l=0$. Each node $i$ in red is placed at an angular position $\theta_i^{(l)}$ on the $\mathbb{S}^1$ circle and has a size proportional to the logarithm of its hidden degree $\kappa_i^{(l)}$. Straight solid lines represent the links in each layer. Coarse-graining blocks correspond to the blue shadowed areas, and dashed lines connect nodes to their supernodes in layer $l+1$. Two supernodes in layer $l+1$ are connected if and only if, in layer $l$, some node in one supernode is connected to some node in the other (blue links give an example).} 
\label{fig1}
\end{figure}

The simplest hidden metric space that can embed a network is a one-dimensional sphere on which nodes have specific angular positions $\{\theta_i ; i=1,\cdots,N\}$. In this space, the transformation proceeds by, first, defining non-overlapping blocks of consecutive nodes of size $r$ along the circle and, second, coarse-graining the blocks into supernodes, regardless of whether they are connected or not to each other. Each supernode is then placed within the angular region defined by the corresponding block so that the order of nodes in the original embedding is preserved in the renormalization process. All the links between some node in one supernode and some node in the other, if any, are renormalized into a single link between the two supernodes. Figure~\ref{fig1} illustrates the process. This coarse-graining procedure is not restricted to equal size blocks and can be defined in different ways as long as the angular distance between the nodes inside the blocks is smaller than the distance between nodes in different blocks. For instance, one could divide the circle in equally sized sectors of a certain arc length such that they contain on average a constant number of nodes. The geometric renormalization operator has abelian semigroup structure with respect to the composition, meaning that a certain number of iterations of a given resolution are equivalent to a single transformation of higher resolution, as shown in Fig.~\ref{fig1}~\footnote{For instance, in Fig.~\ref{fig1} the same transformation with $r = 4$ leads from $l=0$ to $l=2$ in a single step. Whenever the number of nodes is not divisible by $r$, the last supernode in a layer contains less than $r$ nodes, as in the example at $l=1$; however, the RGN equations are valid for uneven supernode sizes as well. Notice that the set of transformations $\mathbb{F}_r$ does not include an inverse element to reverse the process.}. Finally, the set of renormalized network layers $l$, each $r^{l}$ times smaller than the original one, forms a multiscale shell of the network. 

\begin{figure*}[t]
\includegraphics[width=\linewidth]{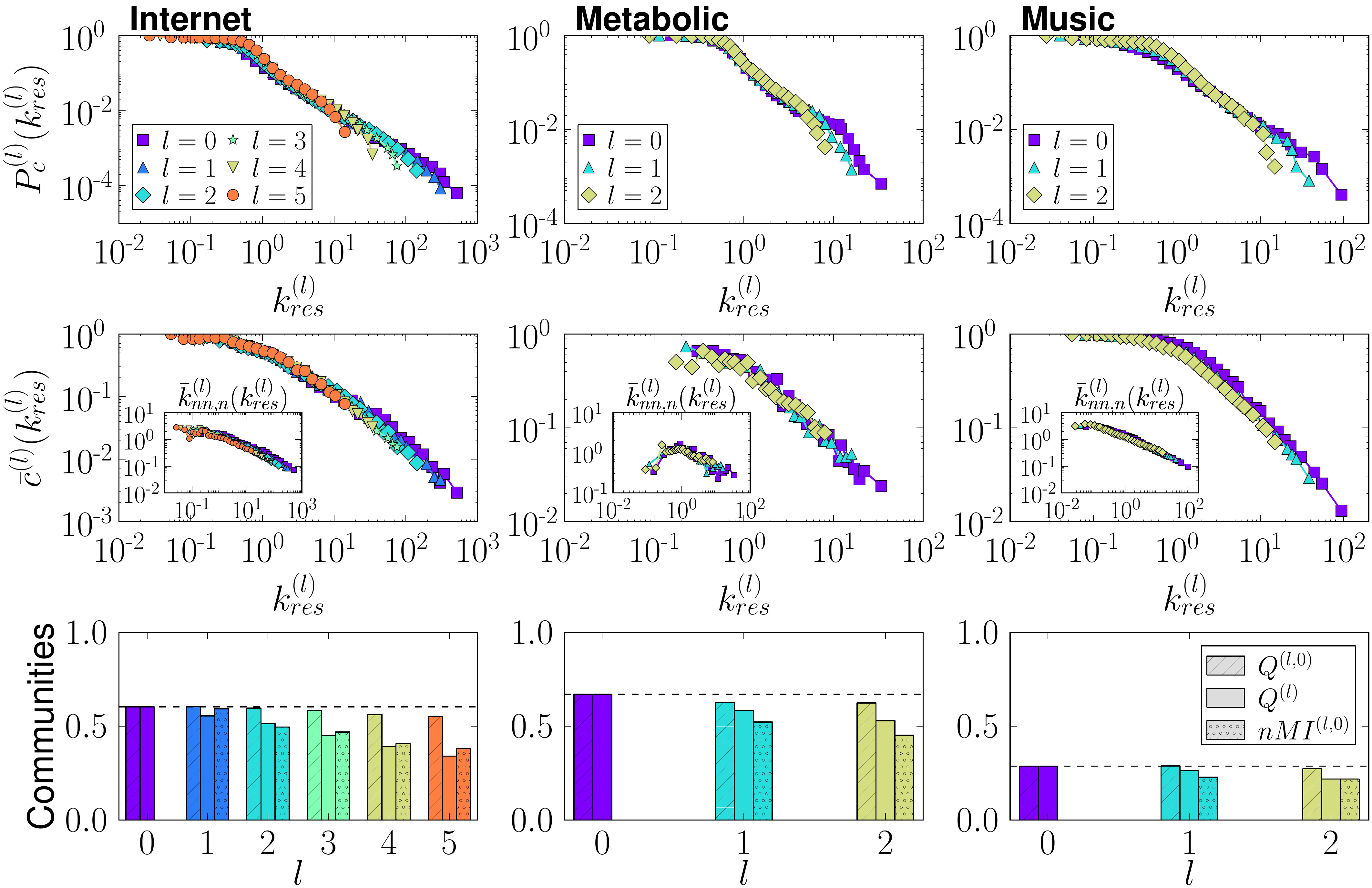}
\caption{{\bf Self-similarity of real networks along the RGN flow.} Each column shows the RGN flow with $r = 2$ of different topological features of the Internet AS network (left), the Human Metabolic network (middle) and the Music network (right).
{\bf Top:} Complementary cumulative distribution of rescaled degrees $k_{res}^{(l)} = k^{(l)}/\langle k^{(l)} \rangle$. {\bf Middle:} Degree-dependent clustering coefficient over rescaled-degree classes. The inset shows the normalized average nearest neighbour degree $\bar{k}_{nn,n} (k_{res}^{(l)}) = \bar{k}_{nn} (k_{res}^{(l)}) \langle k^{(l)} \rangle/\langle (k^{(l)})^2 \rangle$. {\bf Bottom:} RGN flow of the community structure; $Q^{(l)}$ stands for the modularity in layer $l$, $Q^{(l,0)}$ is the modularity that the community structure of layer $l$ induces in the original network, and $nMI^{(l,0)}$ is the normalized mutual information between the latter and the community structure detected directly in the original network. The number of layers in each system is determined by their original size.}
\label{fig2}
\end{figure*}

In this work, we apply the RGN to six different real scale-free networks from very different domains: technology (Internet), transportation (Airports), biology (Cell metabolism and Proteome) and scripts (Music and Words); see Appendix~\ref{app:met} for details. Many real networks can be embedded in the one-dimensional sphere using the $\mathbb{S}^1$ model~\cite{Serrano:2008hb}, which places nodes into a circle and connects every pair with a probability that decreases with their distance along the circle, as a measure of their similarity, and increases with the product of their hidden degrees $\{\kappa_i\}$, as a measure of their popularity (see Appendix~\ref{app:met}). The hidden degrees are well approximated by the observed degrees in the network~\cite{Boguna:2003um,Serrano:2008hb}, and the embedding method uses statistical inference techniques to identify the angular coordinates which maximize the likelihood that the topology of the real network is reproduced by the model~\cite{Boguna2010,frag:hypermap_cn}. Once the hidden degrees and coordinates of the real scale-free networks considered in our study are known, we apply the coarse-graining by defining blocks of size $r=2$ consecutive nodes in the circle, and place the supernodes within the coordinates of their corresponding nodes with the only restriction of preserving the original ordering. We iterate the process so that at each coarse-graining step the size of the system is reduced by a half. 

The resulting topological features of the renormalized networks are shown in Fig.~\ref{fig2} (see also Fig.~\ref{fig:topo} in Appendix~\ref{app:evi}). We observe that the degree distributions, degree-degree correlations ---as measured by the average nearest neighbours degree---, and the clustering spectra, all show self-similar behaviour with curves for the different renormalized layers collapsing if the degrees in the layers are rescaled by their average degree. Also, for every layer $l$ we obtained a partition into communities, $P^{(l)}$, using the Louvain method~\cite{Blondel:2008}; Fig.~\ref{fig2}~bottom shows their modularities $Q^{(l)}$. We also defined the partition induced by $P^{(l)}$ on the original network, $P^{(l,0)}$, obtained by considering that two nodes $i$ and $j$ of the original network are in the same community in $P^{(l,0)}$ if and only if the supernodes of $i$ and $j$ in layer $l$ belong to the same community in $P^{(l)}$. Both the modularity $Q^{(l,0)}$ of $P^{(l,0)}$ and the normalized mutual information $nMI^{(l,0)}$ between both partitions $P^{(0)}$ and $P^{(l,0)}$ are shown in Fig.~\ref{fig2} bottom. Strikingly, the community structure is preserved along the flow to the extent of allowing us to find high-modularity partitions of the original network from much smaller versions of it. This property suggests a new and efficient multiscale community detection algorithm~\cite{Arenas:2008aa,Ronhovde:2009aa,Ahn:2010aa}.

\section{Geometric renormalization of the S$^1$ model}

The self-similarity exhibited by real-world networks can be understood in terms of their congruency with the underlying hidden metric space $\mathbb{S}^1$ model. As we show analytically (see Appendix~\ref{app:rgn} for details), the model is renormalizable in a geometric sense, and that means that real scale-free networks with a geometric structure ---i.~e., which admit a good embedding--- necessarily display the same scaling behaviour.

\begin{figure}[t!]
\centering
\includegraphics[width=\linewidth]{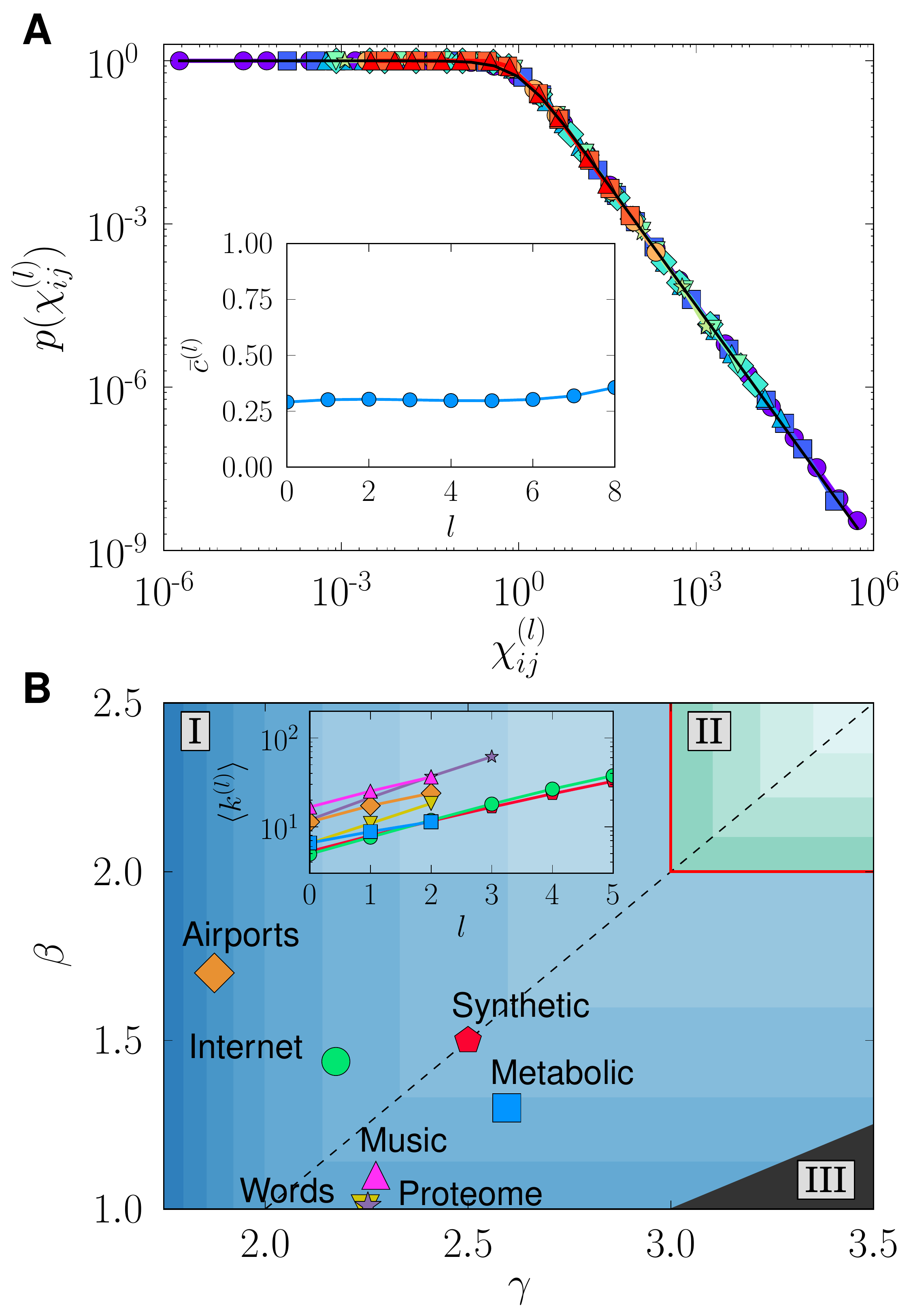}
\caption{ {\bf RGN of the $\mathbb{S}^{1}$ model.}
\textbf{A} Empirical connection probability in a synthetic $\mathbb{S}^1$ network. Fraction of connected pairs of nodes as a function of $\chi_{ij}^{(l)} = R^{(l)} \Delta \theta_{ij}^{(l)}/(\mu^{(l)} \kappa_{i}^{(l)} \kappa_{j}^{(l)})$ in the renormalized layers, from $l=0$ to $l=8$, and $r = 2$. The original synthetic network has $N \sim 225000$ nodes, $\gamma = 2.5$ and $\beta = 1.5$. The black dashed line shows the theoretic curve Eq.~\eqref{eq:conn_prob_S1}. The inset shows the invariance of the mean local clustering along the RGN flow. 
\textbf{B} Real networks in the connectivity phase diagram. The synthetic network above is also shown. Darker blue (green) in the shaded areas represent higher values of the exponent $\nu$. The dashed line separates the $\gamma$-dominated region from the $\beta$-dominated region. In phase I, $\nu > 0$ and the network flows towards a fully connected graph. In phase II, $\nu < 0$ and the network flows towards a one-dimensional ring. The red thick line indicates $\nu = 0$ and, hence, the transition between the small-world and non-small-world phases. In region III, the degree distribution loses its scale-freeness along the flow. The inset shows the exponential increase of the average degree of the renormalized real networks $\langle k^{(l)} \rangle$ with respect to $l$.}
\label{fig3}
\end{figure}

To see why the $\mathbb{S}^1$ model exhibits this self-similarity, we need to consider the renormalization transformation of the geometric layout as well, that is, of hidden degrees, angular positions, $\mu$, $R$ and $\beta$. As we show in Appendix~\ref{app:rgn}, by assigning a new hidden degree $\kappa_i^{(l+1)}$ to supernode $i$ in layer $l+1$ as a function of the hidden degrees of the nodes it contains in layer $l$ according to
\begin{equation}
\kappa_{i}^{(l+1)} = \left( \sum \limits_{j=1}^{r} \left(\kappa_{j}^{(l)}\right)^{\beta} \right)^{1/\beta},
\label{degrees}
\end{equation}
as well as an angular coordinate $\theta_i^{(l+1)}$ given by
\begin{equation}
\theta_{i}^{(l+1)} = \left( \frac{\sum \limits_{j=1}^{r} \left( \theta_{j}^{(l)} \kappa_{j}^{(l)} \right)^{\beta}}{\sum \limits_{j=1}^{r} \left(\kappa_{j}^{(l)}\right)^{\beta}} \right)^{1/\beta},
\label{angles}
\end{equation}
and by rescaling the global parameters as $\mu^{(l+1)} = \mu^{(l)}/r$, $R^{(l+1)} = R^{(l)}/r$ and $\beta^{(l+1)} = \beta^{(l)}$, the renormalized networks remain maximally congruent with the hidden metric space model. This means that the probability $p_{ij}^{(l+1)}$ for two supernodes $i$ and $j$ to be connected in layer $l+1$ (which, according to the RGN procedure is given by the probability for at least one link to exist between some node in $i$ and some node in $j$ in layer $l$), maintains its original form Eq.~\eqref{eq:conn_prob_S1}, as shown in Fig.~\ref{fig3}\textbf{A}. This applies both to the model and to real networks as long as they admit a good embedding, see also Fig.~\ref{fig:conn_prob_regular} in Appendix~\ref{app:evi}. In addition, notice that the transformation of the geometric layout also has the abelian semi-group structure. 

Since the networks remain congruent with the $\mathbb{S}^{1}$ model, hidden degrees $\kappa^{(l)}$ remain proportional to observed degrees $k^{(l)}$, which allows us to explore the degree distribution of the renormalized layers analytically. It can be shown that, if the original distribution of hidden degrees is a power law with characteristic exponent $\gamma$, the hidden degree distribution in the renormalized layers is also a power law with the same exponent asymptotically, as long as $(\gamma-1)/2<\beta$ (see Appendix~\ref{app:rgn}). Interestingly, the global parameter controlling the clustering coefficient, $\beta$, does not change along the flow, which explains the self-similarity of the clustering spectra. Finally, the transformation for the angles Eq.~(\ref{angles}) preserves the ordering of nodes and the heterogeneity in their angular density and, as a consequence, the community structure is preserved in the flow~\cite{Boguna2010,Serrano:2012we,Zuev:2015aa}. The model is therefore renormalizable, and RGN realizations at any scale belong to the same ensemble with a different average degree, which should be rescaled to produce self-similar replicas. 

A good approximation of the behaviour of the average degree for very large networks can be calculated by taking into account the transformation of hidden degrees in the RG flow Eq.~(\ref{degrees}) (see Appendix~\ref{app:rgn} for details). We obtain $\langle k \rangle^{(l+1)} =  r^\nu \langle k \rangle ^{(l)}$, with a scaling factor $\nu$ depending on the connectivity structure of the original network. If $0< \frac{\gamma-1}{\beta} \le 1$, the flow is dominated by the exponent of the degree distribution $\gamma$, and the scaling factor is given by
\begin{equation}\label{eq:approx_avk_upper}
\nu={\frac{2}{\gamma-1} - 1} ,
\end{equation}
whereas the flow is dominated by the strength of clustering if $1 \le \frac{\gamma-1}{\beta} < 2$, and
\begin{equation}\label{eq:approx_avk_lower}
\nu={\frac{2}{\beta} - 1} .
\end{equation}
Therefore, if $\gamma<3$ or $\beta<2$ (phase I in Fig.~\ref{fig3}\textbf{B}), then $\nu>0$ and the model flows towards a highly connected graph; the average degree is preserved if $\gamma = 3$ and $\beta \geq 2$ or $\beta = 2$ and $\gamma \geq 3$, which indicates that the network is at the edge of the transition between the small-world and non-small-world phases; and $\nu<0$ if $\gamma > 3$ and $\beta > 2$, causing the RGN flow to produce sparser networks approaching a unidimensional ring structure as a fixed point (phase II in Fig.~\ref{fig3}\textbf{B}). In this case, the renormalized layers eventually lose the small-world property.

In Fig.~\ref{fig3}\textbf{B}, several real networks are displayed in the connectivity space. All of them lay in the region having the fully connected network as the fixed point, meaning that the RGN flow progressively selects more and more long range connections as a consequence of their small-worldness (see Appendix~\ref{app:rgn}). Furthermore, all of them, except the Internet and the Airports networks, belong to the $\beta$-dominated region. The inset also shows the behaviour of the average degree of every layer $l$, $\langle k^{(l)} \rangle$; as predicted, it grows exponentially in all cases.

Interestingly, global properties of the model, like those reflected in the spectrum of eigenvalues of both the adjacency and laplacian matrices, and quantities like the diffusion time and the restabilization time~\citep{Mieghem:2011}, show a dependence on $\gamma$ and $\beta$ which is in consonance with the one displayed by the RGN flow of the average degree, see results in Figs.~\ref{fig:adjacency}, \ref{fig:laplacian} and \ref{fig:algebraic_connectivity} of Appendix~\ref{app:rgn} for synthetic networks. The $\mathbb{S}^1$ model seems to be more sensitive to small changes in degree heterogeneity in the region $0 < \frac{\gamma-1}{\beta} \le 1$, whereas changes in clustering are better reflected when $1 \le \frac{\gamma-1}{\beta} \le 2$.

\section{Applications}
The RGN enables us to unfold scale-free complex networks in a self-similar multilayer shell which unveils the coexisting scales and their interplay. Beyond the theoretical implications of the discovery that self-similarity under the RGN flow seems to be an ubiquitous symmetry in real networks, their multiscale unfolding can be exploited in immediate practical applications. Next, we propose two among many others; one which singles out a specific scale and another which exploits multiple scales simultaneously. 

\subsection{Mini-me network replicas}
\begin{figure*}[t!]
\includegraphics[width=0.9\linewidth]{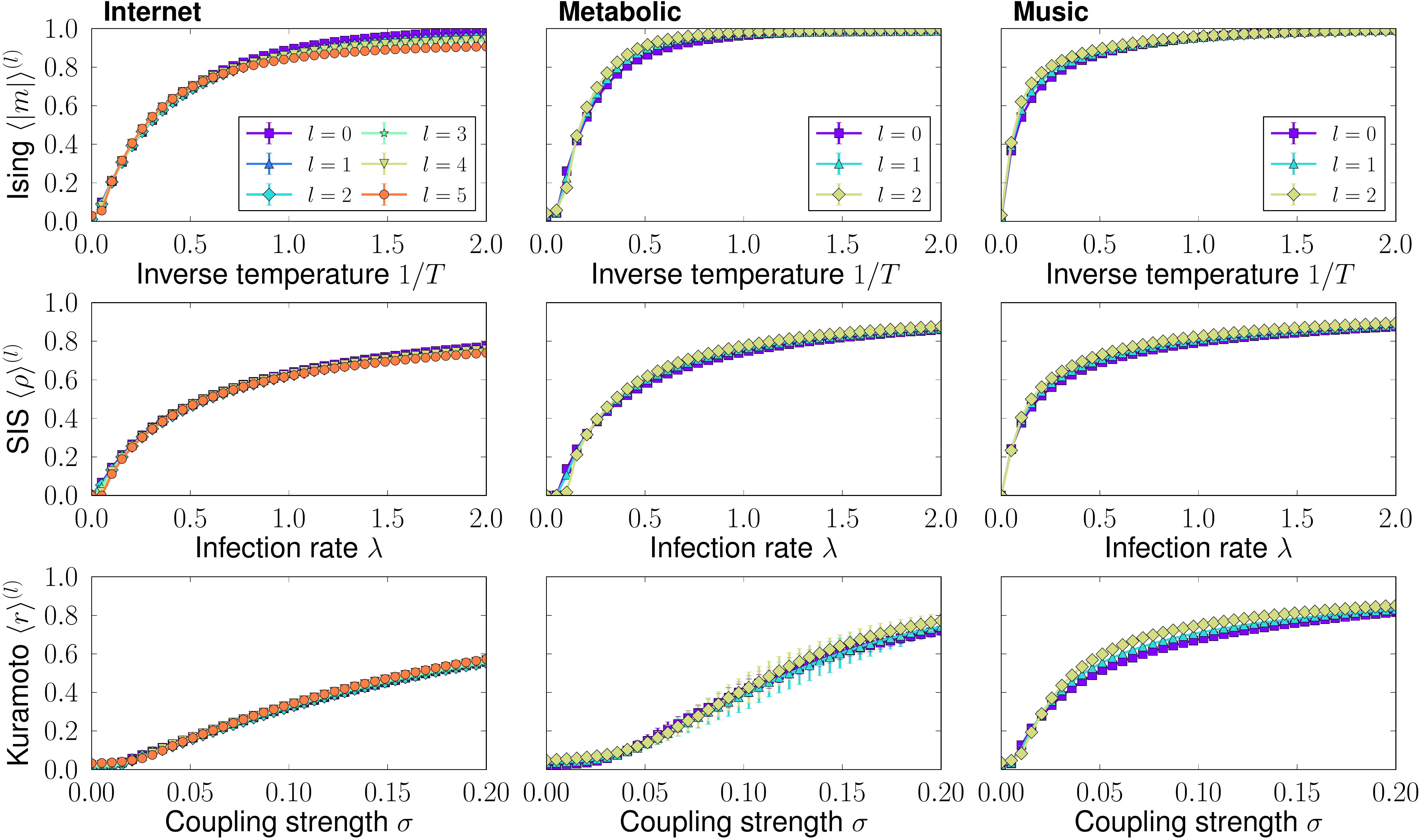}
\caption{{\bf Dynamics on the Mini-me replicas.} Each column shows the order parameters versus the control parameters of different dynamical processes on the original and Mini-me replicas of the Internet AS network (left), the Human Metabolic network (middle) and the Music network (right) with $r = 2$, that is, every value of $l$ identifies a network $2^{l}$ times smaller than the original one. All points show the results averaged over 100 simulations. Error bars indicate the fluctuations of the order parameters. {\bf Top:} Magnetization $\langle | m | \rangle^{(l)}$ of the Ising model as a function of the inverse temperature $1/T$. {\bf Middle:} Prevalence $\langle \rho \rangle^{(l)}$ of the SIS model as a function of the infection rate $\lambda$. {\bf Bottom:} Coherence $\langle r \rangle^{(l)}$ of the Kuramoto model as a function of the coupling strength $\sigma$. In all cases, the curves of the smaller-scale replicas are extremely similar to the results obtained on the original networks.}
\label{fig4}
\end{figure*}

The self-similarity unveiled by the RGN in real networks allows the construction of high-fidelity reduced versions that we call Mini-me network replicas. The downscaling of the topology of large real-world complex networks finds useful applications, for instance, in networked communication systems like the Internet, as a reduced testbed to analyze the performance of new routing protocols~\cite{Papadopoulos:2006aa,Papadopoulos:2007:EIU:1290168.1290173,Yao:2008aa,Yao:2011aa}. However, the success of such program is based upon the quality of the downscaled version of the original network, that should reproduce not only local properties but also the mesoscopic structure of the network. Mini-me replicas can also be used to perform finite size scaling of critical phenomena taking place on real networks, so that critical exponents could be evaluated starting from a single size instance network. The Mini-me networks can be produced at any scale in the range in which self-similarity is preserved. For their construction, we exploit the fact that, under renormalization, a scale-free network remains self-similar and congruent with the underlying geometric model in all the self-similarity range of the multilayer shell. The idea is to single out a specific scale after a certain number of renormalization steps.

Typically, the renormalized average degree of real networks increases in the flow, since they belong to the small-world phase (see inset in Fig.~\ref{fig3}\textbf{B}), meaning that the network layer at the selected scale is more densely connected. To reduce the density to the level of the original network, we apply a pruning of links, see Appendix~\ref{app:met}. Basically, we readjust parameter  $\mu$, controlling the number of links in the underlying geometric $\mathbb{S}^1$ model, so that the expected average degree in the renormalized version is that of the original network, which in turn modifies the connection probability Eq.~(\ref{eq:conn_prob_S1}). We keep in the Mini-me network only the links present in the renormalized layer which are consistent with the readjusted connection probability. In this way, we obtain a reduced version of the real network which is statistically equivalent to a very good approximation. 

To illustrate the high-fidelity that Mini-me network replicas can achieve, we use them to reproduce the behaviour of dynamical processes in real networks. We selected three different dynamical processes, the classic ferromagnetic Ising model, the susceptible-infected-susceptible (SIS) epidemic spreading model, and the Kuramoto model of synchronization, see Appendix~\ref{app:met} for details. We test these dynamics in all the self-similar network layers of the real networks analysed in this work. Results are shown in Fig.~\ref{fig4} and Fig.~\ref{fig:dynamics} in Appendix~\ref{app:dyn}. Quite remarkably, for all dynamics and all networks, we observe very similar results between the original and Mini-me replicas at all scales. This is particularly interesting as these dynamics have a strong dependence on the mesoscale structure of the underlying networks. This strongly supports our claim that both the micro and meso-scales are preserved in the downscaled replicas, as expected given the self-similarity of the network layers in the RGN flow.

\subsection{Multiscale navigation}
Applications that simultaneously exploit more than one or even all the layers in the self-similar multiscale shell are also possible. Next, we introduce a new multiscale navigation protocol for networks embedded in hyperbolic space, which improves single-layer results~\cite{Boguna2010}. To this end, we exploit the quasi-isomorphism between the $\mathbb{S}^1$ model and the $\mathbb{H}^2$ model in hyperbolic space~\cite{Krioukov:2009aa,KrPa10} to produce a purely geometric representation of the multiscale shell (see Appendix~\ref{app:met}). In hyperbolic space, each node is characterised by a radial coordinate directly related to its degree, and an angular coordinate identical to that in the circle. The connection probability becomes a decreasing function of the hyperbolic distance between nodes and, therefore, the most likely path connecting two distant nodes is typically the topological shortest path.

The multiscale protocol is based on greedy routing, in which a source node transmitting information or a packet to a target node sends it to its neighbour closest to destination in the metric space.  As performance metrics we consider the success rate (fraction of successful greedy paths), and the stretch of successful path (ratio between the number of hops in the greedy path and the topological shortest path). Notice that, in general, greedy routing cannot guarantee the existence of a successful greedy path among all pairs of nodes in the network; the packet can get trapped into a loop if sent to an already visited node. In this case, the multiscale protocol can find alternative paths by taking advantage of the increased efficiency of greedy forwarding in the coarse-grained layers. When node $i$ needs to send a packet to some destination node $j$, node $i$ performs a virtual greedy forwarding step in the highest possible layer to find which supernode should be next in the greedy path. Based on this, node $i$ then forwards the packet to its physical neighbour in the real network which guarantees that it will eventually reach such supernode. The process is depicted in Fig.~\ref{fig5}\textbf{A} (full details can be found in Appendix~\ref{app:met}). To guarantee navigation inside supernodes, we require an extra condition in the renormalization process and only consider blocks of connected consecutive nodes. A single node can be left alone forming a supernode by itself, so blocks are of size one or two nodes. Notice that the new requirement does not alter the self-similarity of the renormalized networks forming the multiscale shell (Figs.~\ref{fig:nav_pk} and \ref{fig:nav_ck} in Appendix~\ref{app:mul}) nor the congruency with the hidden metric space (Fig.~\ref{fig:conn_prob_navigation} in Appendix~\ref{app:mul}). 

Figure~\ref{fig5}\textbf{B} shows the increase of the success rate as the number of layers $L$ used in the navigation process is increased for the different real networks considered in this work. Interestingly, as seen in Fig.~\ref{fig5}\textbf{C}, this improvement alters the stretch of successful paths only mildly. The multiscale navigation protocol boosts the success rate by finding paths just slightly longer on average as compared with standard greedy routing in the original network in almost all cases, see inset in Fig.~\ref{fig5}\textbf{C}. The improvement comes at the expense of adding information about the supenodes to the knowledge needed for standard greedy routing in single-layered networks. However, the trade-off between improvement and information overload is advantageous as for many systems the addition of just one or two renormalized layer produces already a notable effect. 

\begin{figure}[t!]
\centering
\includegraphics[width=\linewidth]{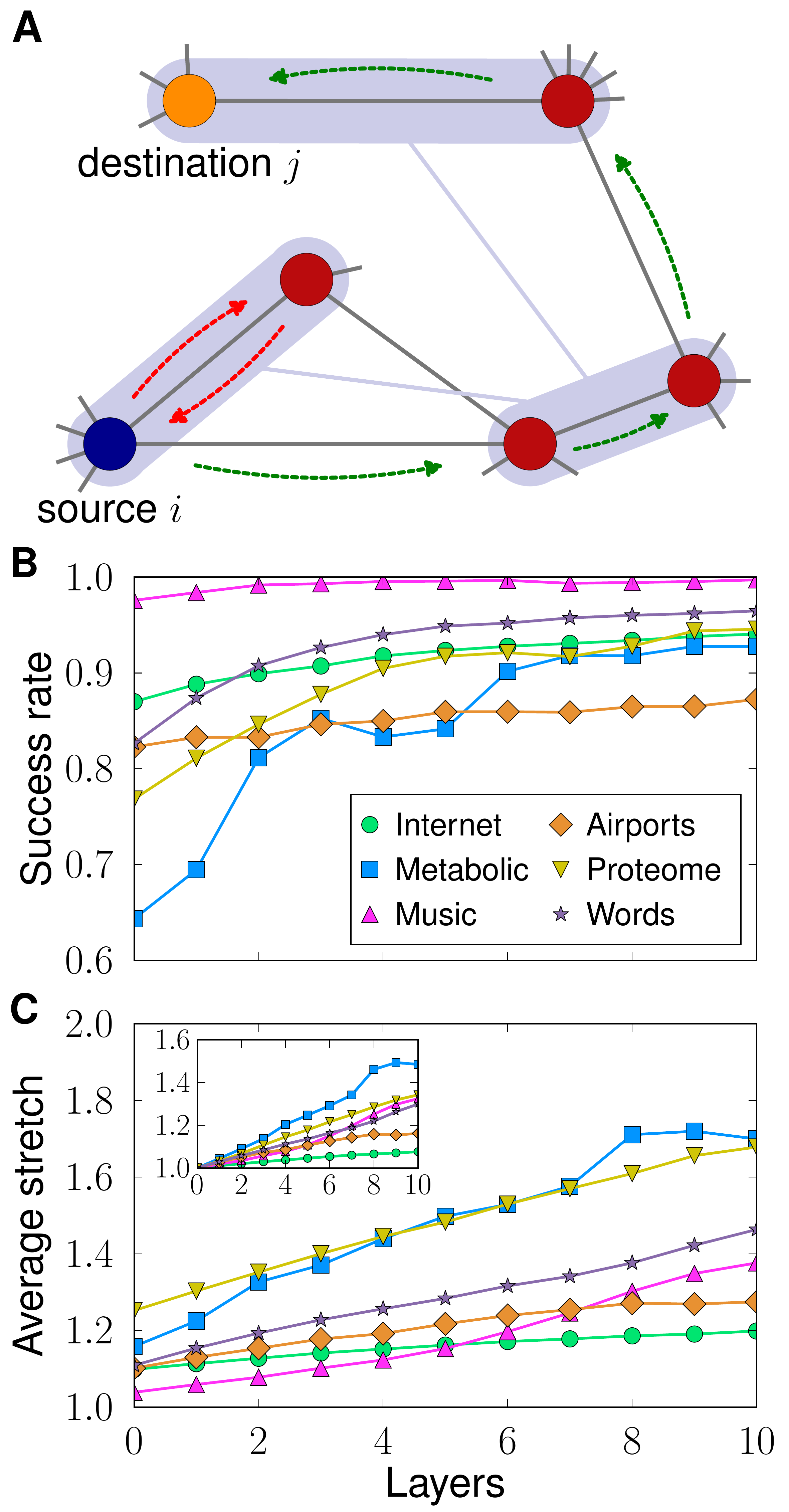}
\caption{ {\bf Multiscale navigation.}
\textbf{A} Illustration of the navigation protocol. Red arrows show the unsuccessful greedy path in the original layer of a message attempting to reach the target yellow node. Green arrows show the successful greedy path from the same source using both layers.
\textbf{B} Success rate as a function of the number of layers used in the process, computed for $10^5$ randomly selected pairs.
\textbf{C} Average stretch $\langle l_{g}/l_{s}\rangle$, where $l_{g}$ is the topological length of a path found by the algorithm and $l_{s}$ is the actual shortest path length in the network. The inset shows the average geometric stretch $\langle l_{g}/l_{g}^{(0)}\rangle$, where $l_{g}^{(0)}$ is the topological length of a path found by the classical single-layer navigation protocol.}
\label{fig5}
\end{figure}

\section{Discussion}

Hidden metric space network models~\cite{Serrano:2008hb,KrPa10,Papadopoulos:2012uq} are able to explain non-trivial structural features of real networks---including scale-free degree distributions, clustering, and self-similarity of the nested hierarchy of subgraphs produced by degree pruning~\cite{Serrano:2011kq}---, and also fundamental mechanisms like preferential attachment in growing networks~\cite{Papadopoulos:2012uq} and the emergence of communities~\cite{Zuev:2015aa}. Interestingly, the existence of a metric space underlying complex networks allows us to define a geometric renormalization group that reveals the multiscale nature of these systems. Quite strikingly, models of scale-free networks are shown to be self-similar under such renormalization, revealing different structural properties depending on the level of coupling with the metric space and degree heterogeneity. The importance of these results, however, stems from the observed self-similarity under geometric renormalization as an ubiquitous symmetry of real world scale-free networks, which moreover stands as a new evidence in favour of the conjecture that hidden metric spaces underlie real networks.

The renormalization group presented in this work is similar in spirit to the topological renormalization studied in~\cite{Song:2005uq}. However, it has clear advantages. First, the ordering in the construction of the boxes is dictated by the embedding of the original network in the underlying space. Second, the congruency between real scale-free networks and the underlying metric space explains the self-similarity of real systems and reveals a multiscale organization that preserves the mesoscopic structure across different observation scales. In the case of topological renormalization, on the other hand, the lack of an underlying model implies that it is not obvious to advance when the network will be self-similar before applying the transformation and whether or not the mesoscopic structure will be mantained. 

From a fundamental point of view, the geometric renormalization group introduced here has proven to be an exceptional tool to unravel the global organization of complex networks across scales and promises to become a standard methodology to analyze real complex networks. It can also help in areas like the study of metapopulation models, in which transportation fluxes or population movements happen both on a local and a global scale~\cite{Colizza:2007fh}. From a practical point of view, we envision many applications besides the two studied in this paper. For instance, the development of a new community detection method that would use the mesoscopic information encoded in the different observation scales, and the use of downscaled versions of the network to perform finite size scaling. This last application would allow for the determination of critical exponents of real complex networks, a task that it not possible with current methods. 

\section*{Acknowledgments}
We acknowledge support from a James S. McDonnell Foundation Scholar Award in Complex Systems; the ICREA Academia prize, funded by the Generalitat de Catalunya; Ministerio de Econom\'{\i}a y Competitividad of Spain projects no. FIS2013-47282-C2-1-P and no. FIS2016-76830-C2-2-P (AEI/FEDER, UE); the Generalitat de Catalunya grant no. 2014SGR608.

\section*{Author contributions}
G.~G.-P., M.~B., and M.~\'A.~S. contributed to the design and implementation of the research, to the analysis of the results, and to the writing of the manuscript.

\section*{Additional information}
{\bf Competing financial interests:} The authors declare no competing financial interests.

\onecolumngrid

\appendix

\section{Methods}\label{app:met}
\subsection{Real networks data}
The real networks analyzed in this paper are:
\begin{itemize}
\item
The Internet at the Autonomous Systems level. The data was collected by the Cooperative Association for Internet Data Analysis (CAIDA)~\citep{Claffy:2009fe} and corresponds to mid 2009.
\item
The Airports network. It was obtained from Ref.~\citep{konect:2016:openflights, konect}. Directed links represent flights by airlines. We consider the undirected version obtained by keeping bidirectional edges only.
\item
The one-mode projection onto metabolites of the human metabolic network at the cell level, as used in Ref.~\citep{Serrano:2012we}. 
\item
The human HI-II-14 interactome. This proteome network was obtained from Ref.~\citep{Rolland20141212}. We removed self-loops.
\item
The Music network. Nodes are chords---sets of musical notes played in a single beat---and connections represent observed transitions among them in a set of songs, see Ref.~\citep{Serra:2012}. The original network is weighted, directed and very dense. Hence, we applied the disparity filter~\citep{Serrano:2009a} with $\alpha = 0.01$ to obtain a sparser network. Finally, we kept bidirectional edges only to construct the undirected network.
\item
The network of adjacency between words in Darwin's book ``On the Origin of Species'', from Ref.~\citep{alon-super-networks}.
\end{itemize}

In all cases, we only considered the largest connected components.

\subsection{$\mathbb{S}^1$ model and transformation to $\mathbb{H}^2$}
The $\mathbb{S}^1$ model~\cite{Serrano:2008hb} places the nodes of a network into a one-dimensional sphere of radius $R$ and connects every pair $i,j$ with probability
\begin{equation}\label{eq:conn_prob_S1}
p_{ij}=\frac{1}{1 + \chi_{ij}^{\beta}}=\frac{1}{1 + \left( \frac{d_{a,ij}}{\mu \kappa_i \kappa_j} \right)^{\beta}}, 
\end{equation}
where $\mu$ controls the average degree of the network, $\beta$ its clustering, and $d_{a,ij}=R \Delta \theta_{ij}$ is the distance between the nodes separated by an angle $\Delta \theta_{ij}$; $R$ is set to $N/2 \pi$, where $N$ is the number of nodes, so that the density of nodes along the circle is equal to 1. The hidden degrees $\kappa_i$ and $\kappa_j$ are proportional to the degrees of nodes $i$ and $j$, respectively. 

The $\mathbb{S}^1$ model is isomorphic to a purely geometric model, the $\mathbb{H}^2$ model~\citep{KrPa10}, in which nodes are placed in a two-dimensional hyperbolic disk of radius
\begin{equation}\label{eq:mapping_Rh2}
R_{\mathbb{H}^2} = 2 \ln \left( \frac{2 R}{\mu \kappa_{0}^{2}} \right),
\end{equation}
where $\kappa_{0} = \min \left\lbrace \kappa_{i} \right \rbrace$. By mapping every mass $\kappa_{i}$ to a radial coordinate $r_{i}$ according to
\begin{equation}\label{eq:mapping_r}
r_{i} = R_{\mathbb{H}^2} - 2 \ln \frac{\kappa_{i}}{\kappa_{0}},
\end{equation}
the connection probability, Eq.~\eqref{eq:conn_prob_S1}, becomes
\begin{equation}\label{eq:pij_h2}
p_{ij} = \frac{1}{1 + e^{\frac{\beta}{2} (x_{ij} - R_{\mathbb{H}^2})}},
\end{equation}
where $x_{ij} = r_{i} + r_{j} + 2 \ln \frac{\Delta \theta_{ij}}{2}$ is a good approximation to the hyperbolic distance between two points with coordinates $(r_{i}, \theta_{i})$ and $(r_{j}, \theta_{j})$ in the native representation of hyperbolic space. The exact hyperbolic distance $d_{\mathbb{H}^2}$ is given by the hyperbolic law of cosines,
\begin{equation}
d_{\mathbb{H}^2} = \mathrm{acosh} \left( \cosh r_i \cosh r_j - \sinh r_i \sinh r_j \cos \Delta \theta_{ij} \right).
\end{equation}

\subsection{Adjusting the average degree of Mini-me network replicas}
To reduce the average degree in a renormalized network to the level of the original network, we apply a pruning of links using the underlying metric model with which the networks in all layers are congruent. The procedure is detailed in this section.

The renormalized network in layer $l$ has an average degree $\langle k^{(l)} \rangle$ generally larger (in phase I) from the original network's $\langle k^{(0)} \rangle$. Moreover, the new network is congruent with the underlying hidden metric space with a parameter $\mu^{(l)} = \mu^{(0)}/r^{l}$ controlling its average degree. The main idea is to decrease the value of $\mu^{(l)}$ to a new value $\mu^{(l)}_{\textrm{new}}$---which implies that the connection probability of every pair of nodes $(i, j)$, $p_{ij}^{(l)}$, decreases to $p_{ij,\textrm{new}}^{(l)}$. We then prune the existing links by keeping them with probability
\begin{equation}\label{eq:qij}
q_{ij}^{(l)} = \frac{p_{ij,\textrm{new}}^{(l)}}{p_{ij}^{(l)}}.
\end{equation}
Therefore, the probability for a link to exist in the pruned network reads,
\begin{equation}
P \lbrace a_{ij,\textrm{new}}^{(l)} = 1 \rbrace = p_{ij}^{(l)} q_{ij}^{(l)} = p_{ij,\textrm{new}}^{(l)},
\end{equation}
whereas the probability for it not to exist is
\begin{equation}
P \lbrace a_{ij,\textrm{new}}^{(l)} = 0 \rbrace = 1 - p_{ij}^{(l)}  + p_{ij}^{(l)} (1 - q_{ij}^{(l)}) = 1 - p_{ij,\textrm{new}}^{(l)},
\end{equation}
that is, the pruned network has a lower average degree and is also congruent with the underlying metric space model with the new value of $\mu^{(l)}_{\textrm{new}}$. Hence, we only need to find the right value of $\mu^{(l)}_{\textrm{new}}$ so that $\langle k^{(l)}_{\textrm{new}} \rangle = \langle k^{(0)} \rangle$. In the thermodynamic limit, the average degree of an $\mathbb{S}^{1}$ network is proportional to $\mu$, so we could simply set
\begin{equation}\label{eq:mu_new}
\mu^{(l)}_{\textrm{new}} = \frac{\langle k^{(0)} \rangle}{\langle k^{(l)} \rangle} \mu^{(l)}.
\end{equation}
However, since we consider real-world networks, finite-size effects play an important role. Indeed, we need to correct the value of $\mu^{(l)}_{\textrm{new}}$ in Eq.~\eqref{eq:mu_new}. To this end, we use a correcting factor $c$, initially set to $c=1$, and use $\mu^{(l)}_{\textrm{new}} = c \frac{\langle k^{(0)} \rangle}{\langle k^{(l)} \rangle} \mu^{(l)}$; for every value of $c$, we prune the network. If $\langle k^{(l)}_{\textrm{new}} \rangle > \langle k^{(0)} \rangle$, we give $c$ the new value $c - 0.1 u \rightarrow c$, where $u$ is a random variable uniformly distributed between 0 and 1. Similarly, if $\langle k^{(l)}_{\textrm{new}} \rangle < \langle k^{(0)} \rangle$, $c + 0.1 u \rightarrow c$. The process ends when $| \langle k^{(l)}_{\textrm{new}} \rangle - \langle k^{(0)} \rangle |$ is below a given threshold (in our case, we set it to 0.1).

\subsection{Simulation of dynamical processes}
The Ising model is an equilibrium model of interacting spins~\cite{dorogovtsev:2002}. Every node $i$ is assigned a variable $s_i$ with two possible values $s_i = \pm 1$, and the energy of the system is, in the absence of external field, given by the Hamiltonian 
\begin{equation}
\mathcal{H} = - \sum\limits_{i<j} J_{ij} a_{ij} s_i s_j,
\end{equation}
where $a_{ij}$ are the elements of the adjacency matrix and $J_{ij}$ are coupling constants which we set to one. We start from an initial condition with $s_i = 1$ for all $i$ and explore the ensemble of configurations using the Metropolis-Hastings algorithm: we randomly select one nod and propose a change in its spin, $- s_i \rightarrow s_i$. If $\Delta \mathcal{H} \leq 0$, we accept the change; otherwise, we accept it with probability $e^{-\Delta \mathcal{H}/T}$, where $T$ is the temperature acting as a control parameter. The order parameter is the absolute magnetization per spin $|m|$, where $m = \frac{1}{N} \sum_{i} s_{i}$; if all spins point in the same direction, $|m| = 1$, whereas $|m| = 0$ if half the spins point in each direction.

In the SIS dynamical model of epidemic spreading~\cite{Pastor-Satorras:2001fl}, every node $i$ can present two states at a given time $t$, susceptible ($n_i(t) = 0$) or infected ($n_i(t) = 1$). Both infection and recovery are Poisson processes. An infected node recovers with rate 1, whereas infected nodes infect their susceptible neighbours at rate $\lambda$. We simulate this process using the continuous-time Gillespie algorithm with all nodes initially infected. The order parameter is the prevalence or fraction of infected nodes $\rho(t) = \frac{1}{N} \sum_{i} n_i(t)$.

The Kuramoto model is a dynamical model for coupled oscillators. Every node $i$ is described by a natural frequency $\omega_i$ and a time-dependent phase $\theta_i(t)$. A node's phase evolves according to
\begin{equation}
\dot{\theta_i} = \omega_i + \sigma \sum\limits_{i<j} a_{ij} \sin( \theta_j(t) - \theta_i(t) ),
\end{equation}
where $a_{ij}$ are the adjacency matrix elements and $\sigma$ is the coupling strength. We integrate the equations of motion using Heun's method. Initially, the phases $\theta_i(0)$  and the frequencies $\omega_i$ are randomly drawn from the uniform distributions $U( -\pi, \pi)$ and $U(-1/2, 1/2)$ respectively, as in Ref.~\citep{pacheco:2004}. The order parameter $r(t) = \frac{1}{N} \left| \sum_j e^{i \theta_j(t)} \right|$ measures the phase coherence of the set of nodes; if all nodes oscillate in phase, $r(t) = 1$, whereas $r(t) \rightarrow 0$ if nodes oscillate in a disordered manner.

In every realization, we compute an average of the order parameter in the stationary state. In the case of the SIS model, the single-realization mean of prevalence values is weighted by time. The curves presented in this work correspond to statistics over 100 realizations.

\subsection{Multiscale navigation}
Given a network and its embedding (layer 0), we merge pairs of consecutive nodes only if they are connected, which guarantees navigation inside supernodes; this process generates layer 1. We repeat the process to generate $L$ layers. The multiscale navigation protocol requires every node $i$ to be provided with the following local information:
\begin{itemize}
\item[1.] The coordinates $(r_i^{(l)}, \theta_i^{(l)})$ of node $i$ in every layer $l$.
\item[2.] The list of (super)neighbours of node $i$ in every layer as well as their coordinates.
\item[3.] Let SuperN$(i,l)$ be the supernode to which $i$ belongs in layer $l$. If SuperN$(i,l)$ is connected to SuperN$(k,l)$ in layer $l$, at least one of the (super)nodes in layer $l-1$ belonging to SuperN$(i,l)$ must be connected to at least one of the (super)nodes in layer $l-1$ belonging to SuperN$(k,l)$; such node is called \textit{gateway}. For every superneighbour of node SuperN$(i,l)$ in layer $l$, node $i$ knows which (super)node or (super)nodes in layer $l-1$ are gateways reaching it. Notice that both the gateways and SuperN$(i, l-1)$ belong to SuperN$(i,l)$ in layer $l$ so, in layer $l-1$, they must either be the same (super)node or different but connected (super)nodes.
\item[4.] If SuperN$(i, l-1)$ is a gateway reaching some supernode $s$, at least one of its (super)neighbours in layer $l-1$ belongs to $s$; node $i$ knows which.
\end{itemize}
This information allows us to navigate the network as follows. Let $j$ be the destination node to which $i$ wants to forward a message, and let node $i$ know $j$'s coordinates in all $L$ layers $(r_j^{(l)}, \theta_j^{(l)})$. In order to decide which of its physical neighbours (i.~e., in layer 0) should be next in the message-forwarding process, node $i$ must first check if it is connected to $j$; in that case, the decision is clear. If it is not, it must:
\begin{itemize}
\item[1.] Find the highest layer $l_{max}$ in which SuperN$(i, l_{max})$ and SuperN$(j, l_{max})$ still have different coordinates. Set $l = l_{max}$.

\item[2.] Perform a standard step of greedy routing in layer $l$: find the closest neighbour of SuperN$(i, l)$ to SuperN$(j, l)$. This is the current target SuperT$(l)$.

\item[3.] While $l>0$, look into layer $l-1$:
\begin{itemize}
\item Set $l = l-1$.
\item If SuperN$(i, l)$ is a gateway connecting to some (super)node within SuperT$(l+1)$, node $i$ sets as new current target SuperT$(l)$ its (super)neighbour belonging to SuperT$(l+1)$ closest to SuperN$(j,l)$.
\item Else node $i$ sets as new target SuperT$(l)$ the gateway in SuperN$(i,l+1)$ connecting to SuperT$(l+1)$ (its (super)neighbor belonging to SuperN$(i,l+1)$).
\end{itemize}

\item[4.] In layer $l=0$, SuperT$(0)$ belongs to the real network and she is a neighbour of $i$, so node $i$ forwards the message to SuperT$(0)$. 
\end{itemize}

\section{Evidence of geometric scaling in real networks}\label{app:evi}
The global topological parameters of all six networks are contained in Table~\ref{tab:networks}.

\begin{table*}[h]
  \begin{ruledtabular}
    \begin{tabular}{llllllll}
\textbf{Name} & \textbf{Type} & \textbf{Nodes} & $N$ & $\gamma$ & $\beta$ & $\langle k \rangle$ & $\langle c \rangle$  \\ \hline
Internet & Technological & Autonomous systems & 23748 & 2.17 & 1.44 & 4.92 & 0.61 \\
Metabolic & Biological & Metabolites & 1436 & 2.6 & 1.3 & 6.57 & 0.54 \\
Music & Script & Chords & 2476 & 2.27 & 1.1 & 16.66 & 0.82 \\
Airports & Transportation & World airports & 3397 & 1.88 & 1.7 & 11.32 & 0.63 \\
Proteome & Biological & Proteins & 4100 & 2.25 & 1.001 & 6.52 & 0.09 \\
Words & Script & Words & 7377 & 2.25 & 1.01 & 11.99 & 0.47 \\
\end{tabular}
\caption{\textbf{Overview of the considered real-world networks.} Details for each dataset can be found in the Appendix~\ref{app:met}.\label{tab:networks}}
\end{ruledtabular}
\end{table*}

\begin{figure}[h!]
\centering
\includegraphics[width=0.9\columnwidth]{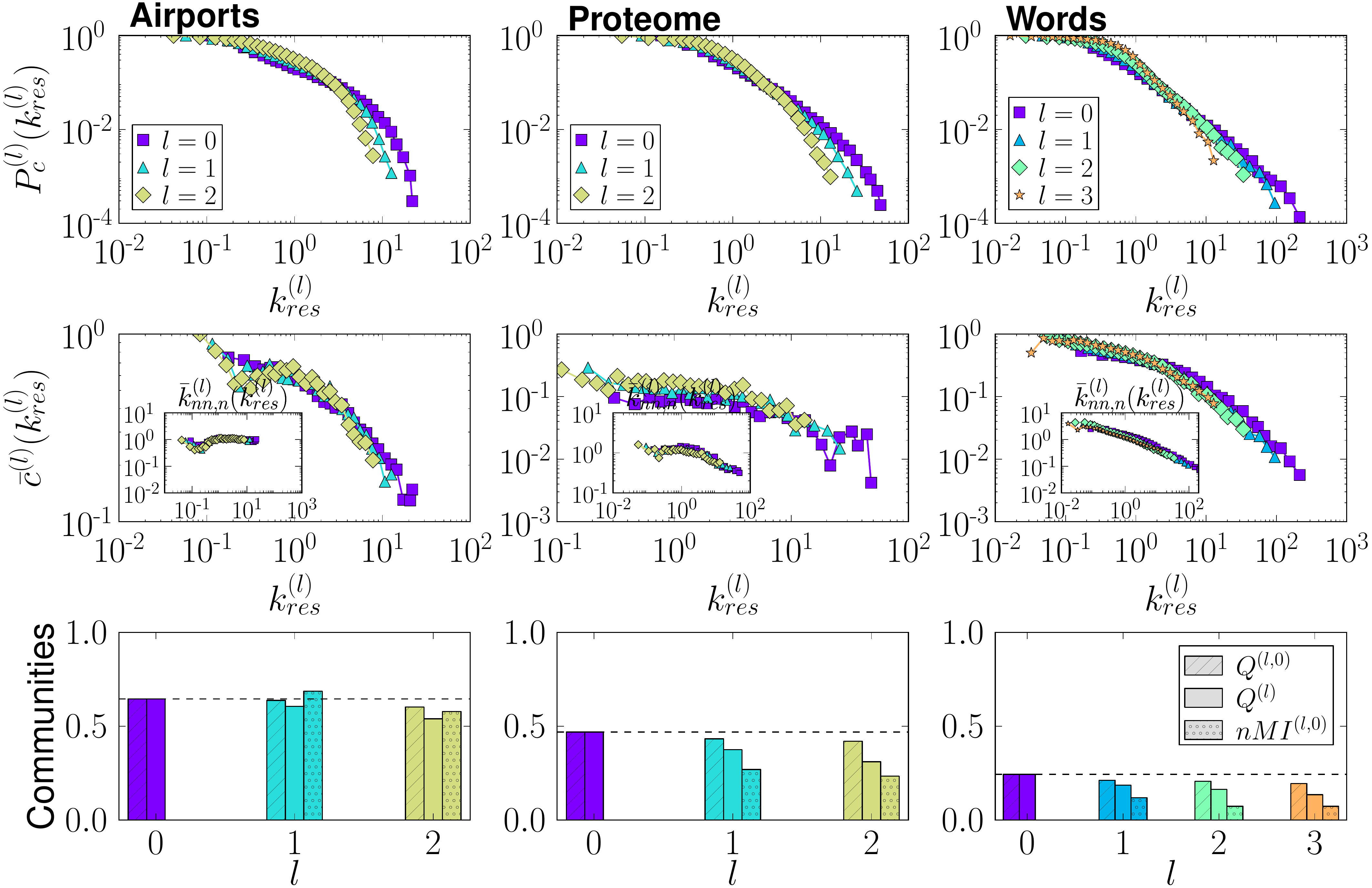}
\caption{{\bf Self-similarity along the RGN flow.} Each column shows the RGN flows of different topological features of the Airports network (left), the Proteome network (middle) and the Words network (right) with $r = 2$.
{\bf Top:} Complementary cumulative distribution of rescaled degrees $k_{res}^{(l)} = k^{(l)}/\langle k^{(l)} \rangle$. {\bf Middle:} Local clustering averaged over rescaled-degree classes. The inset shows the normalized average nearest neighbour degree $\bar{k}_{nn,n} (k_{res}^{(l)}) = \bar{k}_{nn} (k_{res}^{(l)}) \langle k^{(l)} \rangle/\langle (k^{(l)})^2 \rangle$. {\bf Bottom:} RGN flow of the community structure; $Q^{(l)}$ stands for the modularities in every layer $l$, $Q^{(l,0)}$ is the modularity that the community structure in the $l$ layer induces in the original network, and $nMI^{(l,0)}$ is the normalized mutual information between both partitions. The number of layers in each system is determined by their original size.}
\label{fig:topo}
\end{figure}

Fig.~\ref{fig2} compares the topological properties of the renormalized networks for three real networks. We show the equivalent results for the Airports, Proteome and Words networks in Fig.~\ref{fig:topo}.

\begin{figure}[h!]
\centering
\includegraphics[width=1.0\columnwidth]{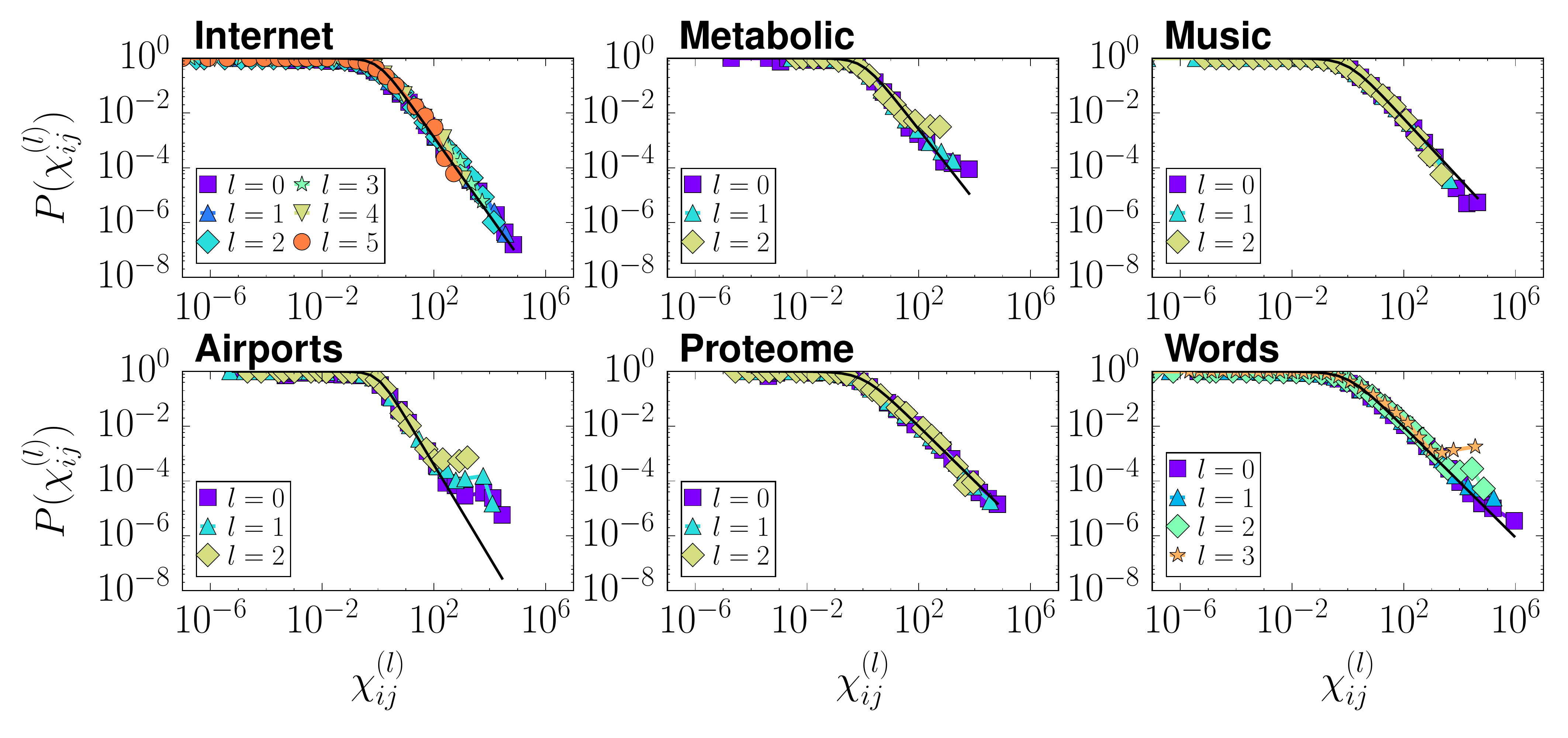}
\caption{\label{fig:conn_prob_regular} \textbf{Empirical connection probabilities.} Fraction of connected pairs within a given range of $\chi_{ij}^{(l)}$ for the six real-world networks and their renormalized versions. The black curve is the theoretic connection probability.}
\end{figure}

In Fig.~\ref{fig:conn_prob_regular}, we show the empirical connection probabilities of the six real-world networks considered in this paper as well as their renormalized versions. 

\section{The Geometric Renormalization Group}\label{app:rgn}
This section contains the calculations related to the theoretical aspects of the geometric renormalization transformation. In particular, we show the semi-group structure of the transformation, derive the corresponding recurrence relations for the renormalization of the $\mathbb{S}^1$ model and calculate the flow of the average degree. We also discuss the connection with statistical mechanics by using the isomorphism between the $\mathbb{S}^1$ and the $\mathbb{H}^2$ models and, finally, we include some numerical results regarding the relation between global properties of the networks generated by the model and the flow of the average degree.

\subsection{The semigroup structure of the coarse-graining step}
It is easy to show that the geometric coarse-graining presented in this paper has the semigroup structure. To this end, we need to see that node $i$ is mapped to the same supernode whether we apply the coarse-graining with $r = r_{1}$ first and then a second time with $r = r_{2}$ or just once with $r = r_{1}r_{2}$. In the first case, the step with $r = r_{1}$ maps $i$ to supernode $m = \left\lfloor i/r_{1} \right\rfloor$ (where $\lfloor x \rfloor$ represents the integer part of $x$), and then $m$ is mapped to $n = \left\lfloor m/r_{2} \right\rfloor = \left\lfloor \left\lfloor i/r_{1} \right\rfloor/r_{2} \right\rfloor$ in the second step. In the second case, $i$ is mapped to supernode $s = \left\lfloor i/(r_{1}r_{2}) \right\rfloor$. Notice that $s = \left\lfloor ( i/r_{1} )/r_{2} \right\rfloor = \left\lfloor ( \left\lfloor i/r_{1} \right\rfloor + \alpha)/r_{2} \right\rfloor = \left\lfloor ( m + \alpha)/r_{2} \right\rfloor$, where $\alpha = (i \mod r_{1})/r_{1} < 1$. Now,
\begin{equation}
\frac{m+\alpha}{r_2} = \frac{m}{r_2} + \frac{\alpha}{r_2} = \left \lfloor \frac{m}{r_2} \right \rfloor + \frac{m\mod r_2}{r_2} + \frac{\alpha}{r_2},
\end{equation}
so
\begin{equation}
\left \lfloor \frac{m+\alpha}{r_2} \right \rfloor = \left \lfloor \frac{m}{r_2} \right \rfloor \Leftrightarrow \frac{m \mod r_2}{r_2} + \frac{\alpha}{r_2} < 1,
\end{equation}
which is always fulfilled since $\alpha < 1$ and
\begin{equation}
\frac{m \mod r_2 + \alpha}{r_2} \leq \frac{r_2 - 1 + \alpha}{r_2} = 1 + \frac{\alpha - 1}{r_2} < 1.
\end{equation}
Thus, $s = n$, and node $i$ is mapped to the same supernode in both cases. It follows immediately from this result that both processes yield the same final link structure.

\subsection{Selecting long-range connections}
\begin{figure}[h!]
\centering
\includegraphics[width=0.5\columnwidth]{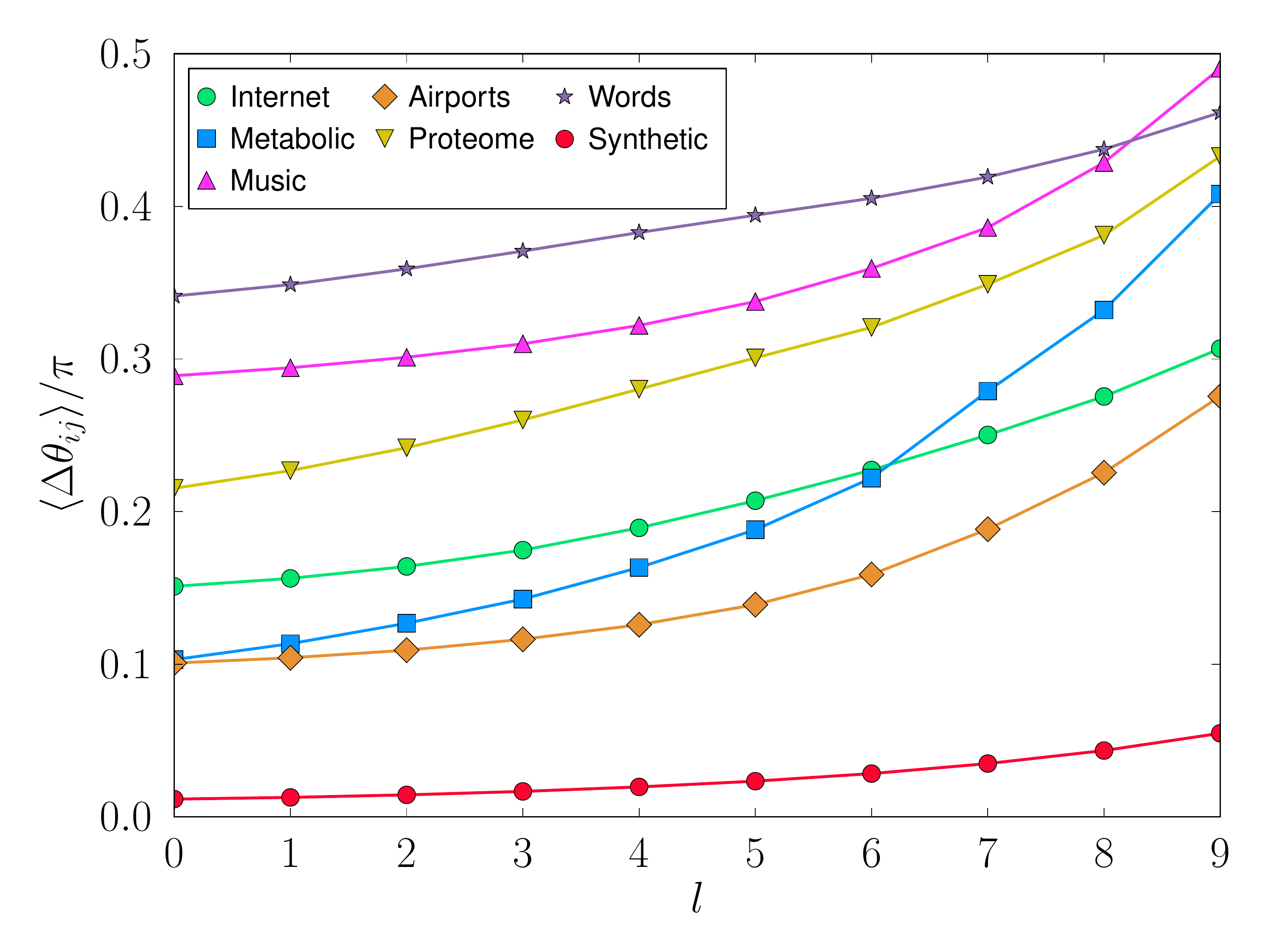}
\caption{\label{fig:range}\textbf{Connection range in renormalized layers.} Normalized angular distance $\langle \Delta \theta_{ij} \rangle$ averaged over all links that are not integrated inside a supernode in layer $l$ with $r=2$.}
\end{figure}

As we apply the renormalization transformation, some links are integrated inside the supernodes, so they do not contribute to the topology of the renormalized network. In Fig.~\ref{fig:range}, we show that links joining nodes separated a large angular distance $\Delta \theta_{ij}$ require larger values of $r$ to be integrated; in other words, the connections in a renormalized network represent long-range connections in the original graph.

\subsection{Geometric renormalization of the S$^1$ model}

In this subsection, we derive the RG equations of the $\mathbb{S}^1$ model. In order to simplify the notation, all unprimed quantities will refer to layer $l-1$, whereas primed ones will correspond to layer $l$. Moreover, we consider the particular case in which all supernodes contain the same number of nodes ($r$) for simplicity, although the following calculations are also valid for supernodes of different sizes.

Consider the probability $p_{ij}'$ for two supernodes $i$ and $j$ in layer $l$ to be connected, which is given by the probability for at least one link between a pair of the nodes within the supernodes in layer $l-1$ to exist,
\begin{equation}\label{eq:pij_primed}
p_{ij}' = 1 - \prod \limits_{e = 1}^{r^2} \left( 1 - p_{e} \right),
\end{equation}
where $e$ runs over all pairs of nodes $(m,n)$ with $m$ in supernode $i$ and $n$ in supernode $j$. The term $p_{e}$ is the probability for $m$ and $n$ to be connected in layer $l-1$,
\begin{equation}\label{eq:pij_original}
p_{e} = \frac{1}{1 + \left( \frac{R \Delta \theta_{e}}{\mu (\kappa_{m} \kappa_{n})_{e}} \right)^{\beta}}.
\end{equation}
Eq.~\eqref{eq:pij_primed} takes the same functional form as Eq.~\eqref{eq:pij_original},
\begin{equation}\label{eq:pij_fermi}
p_{ij}' = 1 - \prod \limits_{e = 1}^{r^2} \frac{1}{1 + \left( \frac{R \Delta \theta_{e}}{\mu (\kappa_{m} \kappa_{n})_{e}} \right)^{- \beta}} 
= 1 - \frac{1}{\prod \limits_{e = 1}^{r^2} 1 + \left( \frac{\mu (\kappa_{m} \kappa_{n})_{e}}{R \Delta \theta_{e}} \right)^{ \beta}}
= 1 - \frac{1}{1 + \Phi_{ij}'} = \frac{1}{1 + \frac{1}{\Phi_{ij}'}}
\end{equation}
with
\begin{equation}\label{eq:phi_exact}
\Phi_{ij}' = \sum \limits_{e = 1}^{r^2} \left( \frac{\mu (\kappa_{m} \kappa_{n})_{e}}{R \Delta \theta_{e}} \right)^{\beta} + \sum \limits_{e = 1}^{r^2 - 1} \sum \limits_{f = e + 1}^{r^2} \left( \frac{\mu (\kappa_{m} \kappa_{n})_{e}}{R \Delta \theta_{e}} \right)^{\beta} \left( \frac{\mu (\kappa_{m} \kappa_{n})_{f}}{R \Delta \theta_{f}} \right)^{\beta} + \ldots
\end{equation}
Since the angular distance between the nodes inside each block is generally smaller than the distance between $i$ and $j$, all the $\Delta \theta_{e}$ are approximately equal ($\Delta \theta_e \approx \Delta \theta$), so we can write
\begin{equation}\label{eq:phi_expansion_const_angle}
\Phi_{ij}' \approx \left( \frac{\mu}{R \Delta \theta} \right)^{\beta} \sum \limits_{e = 1}^{r^2} (\kappa_{m} \kappa_{n})_{e}^{\beta} + \left( \frac{\mu}{R \Delta \theta} \right)^{2 \beta} \sum \limits_{e = 1}^{r^2 - 1} \sum \limits_{f = e + 1}^{r^2} (\kappa_{m} \kappa_{n})_{e}^{\beta} (\kappa_{m} \kappa_{n})_{f}^{\beta} + \ldots
\end{equation}
The $\mathbb{S}^1$ model assumes a uniform density of nodes $\delta = 1$, which means that $R = \frac{N}{2 \pi}$, whereas $\mu$ is a constant independent of $N$. Indeed, $\frac{\mu}{R} \ll 1$, so the first term leads Eq.~\eqref{eq:phi_expansion_const_angle} in most cases. Thus,
\begin{equation}\label{eq:phi_approx}
\Phi_{ij}' \approx \left( \frac{\mu}{R \Delta \theta} \right)^{\beta} \sum \limits_{e = 1}^{r^2} (\kappa_{m} \kappa_{n})_{e}^{\beta}.
\end{equation}
Introducing this result into Eq.~\eqref{eq:pij_fermi},
\begin{equation}
p_{ij}' \approx \frac{1}{1 + \left( \frac{R \Delta \theta}{\mu} \right)^{\beta} \frac{1}{\sum \limits_{e = 1}^{r^2} (\kappa_{m} \kappa_{n})_{e}^{\beta}}},
\end{equation}
we see that, in order for the resulting expression to be congruent with the model, we need a set of equations that transform the parameters according to
\begin{equation}\label{eq:params_transf}
\left( \frac{R \Delta \theta}{\mu} \right)^{\beta} \frac{1}{\sum \limits_{e = 1}^{r^2} (\kappa_{m} \kappa_{n})_{e}^{\beta}} = \left( \frac{R' \Delta \theta_{ij}'}{\mu' \kappa'_{i} \kappa'_{j}} \right)^{\beta'}.
\end{equation}
Let us now assume that the angular coordinate of a supernode is some generalised center of mass of the nodes it integrates, so the separation between the two renormalised nodes $\Delta \theta_{ij}'$ is approximately equal to the angular separation between the nodes that belong to different blocks, i.e. $\Delta \theta_{ij}' \approx \Delta \theta$; thus, $\beta' = \beta$. The choice $\delta = 1$ leads to $R' = \frac{R}{r}$, that is, to the rescaling step. Setting $\mu' = \frac{\mu}{r}$, Eq.~\eqref{eq:params_transf} further requires
\begin{equation}
\left( \kappa'_{i} \kappa'_{j} \right)^{\beta} = \sum \limits_{e = 1}^{r^2} (\kappa_{m} \kappa_{n})_{e}^{\beta},
\end{equation}
which is fulfilled if
\begin{equation}\label{eq:kappa_p}
\kappa'_{i} = \left( \sum \limits_{j=1}^{r} \kappa_{j}^{\beta} \right)^{1/\beta}.
\end{equation}
The transformation of masses preserves the semi-group structure exactly, since
\begin{equation}\label{eq:theta_p}
\left(\kappa''_{i}\right)_{r} = \left( \sum \limits_{j=1}^{r} \left(\kappa_{j}'\right)^{\beta} \right)^{1/\beta} = \left( \sum \limits_{j=1}^{r} \sum \limits_{k=1}^{r} \kappa_{j,k}^{\beta} \right)^{1/\beta} = \left(\kappa'_{i}\right)_{r^2}.
\end{equation}
We should require the transformation of angles to preserve it as well. This can be achieved using the following generalised center of mass
\begin{equation}
\theta'_{i} = \left( \frac{\sum \limits_{j=1}^{r} \left( \theta_{j} \kappa_{j} \right)^{\beta}}{\sum \limits_{j=1}^{r} \kappa_{j}^{\beta}} \right)^{1/\beta},
\end{equation}
given that
\begin{equation}
\left(\theta''_{i}\right)_{r} = \frac{1}{\left(\kappa''_{i}\right)_{r}} \left( \sum \limits_{j=1}^{r} \left( \theta'_{j} \kappa'_{j} \right)^{\beta} \right)^{1/\beta} = \frac{1}{\left(\kappa'_{i}\right)_{r^2}} \left( \sum \limits_{j=1}^{r} \left(\kappa_{j}'\right)^{\beta} \frac{1}{\left(\kappa_{j}'\right)^{\beta}} \sum \limits_{k=1}^{r} \left( \theta_{j,k} \kappa_{j,k} \right)^{\beta} \right)^{1/\beta} = \left(\theta'_{i}\right)_{r^2}.
\end{equation}

\subsection{RG flow of the average degree}
As discussed in the previous subsection, as we renormalize, we move in the space of realizations of the $\mathbb{S}^1$ model, always keeping the congruency between the network and the hidden metric space, i.e.~Eq.~\eqref{eq:pij_original}. Therefore, we can use the $\mathbb{S}^1$ model to compute the average degree $\langle k' \rangle$ of the renormalised networks. According to Ref.~\citep{KrPa10},
\begin{equation}\label{eq:av_k_av_kappa}
\langle k' \rangle = C_0 \mu' \langle \kappa' \rangle^2,
\end{equation}
where $C_0$ does not change as we renormalize. We thus need to compute $\langle \kappa' \rangle$, where $\kappa'$ is given by Eq.~\eqref{eq:kappa_p} and the original distribution of masses is assumed to be a power-law,
\begin{equation}\label{eq:kappa_dist}
\rho(\kappa) = \frac{1-\gamma}{\kappa_c^{1 - \gamma}-\kappa_0^{1 - \gamma}} \kappa^{- \gamma}, \quad \kappa \in [\kappa_0, \kappa_c].
\end{equation}
The strategy to compute $\langle \kappa' \rangle$ is as follows: \textbf{1.}~We define $z \equiv \kappa^\beta$ and find their distribution $\rho_z (z)$. \textbf{2.}~We then calculate $\hat{\rho}_z^r(s)$ (where $\hat{\rho}_z(s)$ is the Laplace transform of $\rho_z(z)$); according to the convolution theorem, this is the Laplace transform of the variable $z' \equiv \sum_r z = \kappa'^\beta$. \textbf{3.}~Finally, we compute $\langle \kappa' \rangle$ as the $1/\beta$-th moment of $z'$, that is, $\langle \kappa' \rangle = \langle z'^{1/\beta} \rangle$, from $\hat{\rho}_z^r(s)$.
\begin{itemize}

\item[\textbf{1.}] From Eq.~\eqref{eq:kappa_dist},
\begin{equation}
\rho(\kappa) \text{d}k = \frac{1-\gamma}{\kappa_c^{1 - \gamma}-\kappa_0^{1 - \gamma}} \kappa^{- \gamma} \text{d}\kappa =  \frac{1-\gamma}{\beta \left(\kappa_c^{1 - \gamma}-\kappa_0^{1 - \gamma}\right)} z^{\frac{1-\gamma}{\beta} -1} \text{d}z,
\end{equation}
so
\begin{equation}\label{eq:rho_z}
\rho_z(z) = \frac{1-\gamma}{\beta \left(\kappa_c^{1 - \gamma}-\kappa_0^{1 - \gamma}\right)} z^{- \eta},
\end{equation}
where $\eta = \frac{\gamma-1}{\beta} + 1$.

\item[\textbf{2.}] If $\gamma < 2 \beta + 1$, $\eta < 3$, which means that $z'$ and, consequently, $\kappa'$ are also power-law distributed since the central limit theorem does not apply (the opposite case corresponds to phase III in Fig.2{\bf B})~\citep{gnedenko}. The Laplace transform of Eq.~\eqref{eq:rho_z} is given by
\begin{equation}\label{eq:rho_hat_z}
\hat{\rho}_z(s) = \int \limits_{\kappa_0^\beta}^{\kappa_c^\beta} \rho_z (z) e^{- s z} \text{d}z = \frac{\left(1-\gamma\right)\left( \Gamma(1-\eta, s \kappa_0^\beta) - \Gamma(1-\eta, s \kappa_c^\beta) \right)}{\beta \left(\kappa_c^{1 - \gamma}-\kappa_0^{1 - \gamma}\right)} s^{\eta - 1},
\end{equation}
where $\Gamma(a,b)$ is the incomplete gamma function,
\begin{equation}
\Gamma(a,b) = \int \limits_b^\infty t^{a-1}e^{-t} \text{d}t.
\end{equation}
From this result, it follows that
\begin{equation}\label{eq:rho_hat_zp}
\hat{\rho}_{z'}(s) = \left[ \frac{\left(1-\gamma\right)\left( \Gamma(1-\eta, s \kappa_0^\beta) - \Gamma(1-\eta, s \kappa_c^\beta) \right)}{\beta \left(\kappa_c^{1 - \gamma}-\kappa_0^{1 - \gamma}\right)} s^{\eta - 1} \right]^r.
\end{equation}

\item[\textbf{3.}] We need to compute
\begin{equation}
\langle z'^{1/\beta} \rangle = \int \limits_0^\infty z'^{1/\beta} \rho_{z'}(z') \text{d}z'.
\end{equation}
To do so, consider the integral
\begin{equation}
I = C' \int \limits_0^\infty s^\alpha \hat{\rho}_{z'}^{(n)}(s) \text{d}s = C' \int \limits_0^\infty s^\alpha \int \limits_0^\infty (-1)^n z'^n \rho_{z'}(z') e^{-sz'} \text{d} z' \text{d} s.
\end{equation}
Taking into account that for $\alpha > -1$
\begin{equation}
\int \limits_0^\infty s^\alpha e^{-sz'} \text{d} s = z'^{-1-\alpha} \Gamma(1+\alpha),
\end{equation}
we see that
\begin{equation}
I = C' (-1)^n \Gamma(1+\alpha) \int \limits_0^\infty z'^{n-1-\alpha} \rho_{z'}(z') \text{d} z'.
\end{equation}
Now, setting $C' = (-1)^n \Gamma(1+\alpha)^{-1}$ and $n-1-\alpha = 1/\beta$, $I = \langle z'^{1/\beta} \rangle$. However, since $\alpha = n - 1 - 1/\beta > -1$ and $n \in \mathbb{N}$, the smallest $n$ we can choose is $n = 1$, so $\alpha = -1/\beta$. Finally, we can write
\begin{equation}\label{eq:av_kappap}
\langle \kappa' \rangle = -\frac{1}{\Gamma \left(1- \frac{1}{\beta}\right)} \int \limits_0^\infty s^{-1/\beta} \hat{\rho}_{z'}'(s) \text{d}s,
\end{equation}
where $\hat{\rho}_{z'}(s)$ is given in Eq.~\eqref{eq:rho_hat_zp}.
\end{itemize}

\subsubsection*{Particular case $r=2$}
To start solving Eq.~\eqref{eq:av_kappap}, let us first take the limit of $N \to \infty \Rightarrow \kappa_c \to \infty$, which means that $\hat{\rho}_z(s)$ becomes
\begin{equation}\label{eq:rho_hat_z_c}
\hat{\rho}_z(s) = C s^{\eta - 1} \Gamma(1-\eta, s \kappa_0^\beta), \quad C = \frac{\gamma - 1 }{\beta \kappa_0^{1 - \gamma}}.
\end{equation}
Using the same change of variable as in Eq.~\eqref{eq:rho_hat_z}, we see that
\begin{equation}
\hat{\rho}_z'(s) = - \int \limits_{\kappa_0^\beta}^{\kappa_c^\beta} z \rho_z (z) e^{- s z} \text{d}z = - C s^{\eta - 2} \Gamma(2-\eta, s \kappa_0^\beta).
\end{equation}
Let us now evaluate $\hat{\rho}_{z'}'(s) $,
\begin{equation}
\hat{\rho}_{z'}'(s) = r \hat{\rho}_{z}^{r-1}(s) \hat{\rho}_{z}'(s) = 2 \hat{\rho}_{z}(s) \hat{\rho}_{z}'(s) = - 2 C^2 s^{2 \eta - 3}\Gamma(1-\eta, s \kappa_0^\beta)\Gamma(2-\eta, s \kappa_0^\beta)
\end{equation}
and introduce this result into Eq.~\eqref{eq:av_kappap},
\begin{equation}\label{eq:av_kappap_b2}
\begin{aligned}
\langle \kappa' \rangle &= \frac{2 C^2}{\Gamma \left(1- \frac{1}{\beta}\right)} \int \limits_0^\infty s^{2 \eta - 3 - 1/\beta} \Gamma(1-\eta, s \kappa_0^\beta)\Gamma(2-\eta, s \kappa_0^\beta) \text{d}s \\
&= \frac{2 C^2}{\Gamma \left(1- \frac{1}{\beta}\right)} \frac{1}{\kappa_0^{\beta(2 \eta - 2 - 1/\beta)}} \int \limits_0^\infty \omega^{2 \eta - 3 - 1/\beta} \Gamma(1-\eta, \omega)\Gamma(2-\eta, \omega) \text{d}\omega \\
&=  \kappa_0 \frac{2 (\gamma - 1)^2}{\beta^2 \Gamma \left(1- \frac{1}{\beta}\right)} \int \limits_0^\infty \omega^{2 \eta - 3 - 1/\beta} \Gamma(1-\eta, \omega)\Gamma(2-\eta, \omega) \text{d}\omega.
\end{aligned}
\end{equation}
We thus need to solve an integral of the form
\begin{equation}
I(\nu, s_1, s_2) = \int \limits_0^\infty x^{\nu} \Gamma(s_1,x) \Gamma(s_2,x) \text{d}x, \quad \nu > -1.
\end{equation}
In our case, $\nu = 2 \eta - 3 - 1/\beta = (2 \gamma - 3)/\beta - 1 > -1 \Leftrightarrow \gamma > 3/2$. Integrating by parts,
\begin{equation}
\begin{aligned}\label{eq:i_nu_s1_s2}
I(\nu, s_1, s_2) &= \left. \frac{1}{\nu + 1} x^{\nu + 1} \Gamma(s_1,x) \Gamma(s_2,x) \right|_0^\infty + \frac{1}{\nu+1} \int \limits_0^\infty x^{\nu+1} \left( \Gamma(s_1,x) x^{s_2-1}e^{-x} +  \Gamma(s_2,x) x^{s_1-1}e^{-x} \right) \text{d}x \\
&= \frac{1}{\nu+1} \int \limits_0^\infty \left( \Gamma(s_1,x) x^{\nu + s_2}e^{-x} +  \Gamma(s_2,x) x^{\nu + s_1}e^{-x} \right) \text{d}x.
\end{aligned}
\end{equation}
We can find a recurrence relation for the integrals in the last expression,
\begin{equation}
\begin{aligned}
I'(\alpha, s) &=  \int \limits_0^\infty \Gamma(s,x) x^{\alpha}e^{-x} \text{d}x = \left. \frac{1}{\alpha + 1} x^{\alpha + 1} \Gamma(s,x) \right|_0^\infty + \frac{1}{\alpha + 1} \int \limits_0^\infty x^{\alpha+1} \left( x^{s-1} e^{-2x} + \Gamma(s,x) e^{-x} \right) \text{d}x \\
&= \left. \frac{1}{\alpha + 1} \frac{1}{2^{\alpha+s+1}}\Gamma(\alpha + s + 1, 2x) \right|_0^\infty + \frac{1}{\alpha + 1} \int \limits_0^\infty \Gamma(s,x) x^{\alpha + 1}e^{-x} \text{d}x \\
&= \frac{1}{\alpha + 1} \frac{1}{2^{\alpha+s+1}}\Gamma(\alpha + s + 1) + \frac{1}{\alpha + 1} I'(\alpha+1,s).
\end{aligned}
\end{equation}
Iterating yields
\begin{equation}
I'(\alpha, s) = \sum \limits_{n=1}^{\infty} \frac{1}{\prod \limits_{n'=1}^n (\alpha + n')}\frac{1}{2^{\alpha+s+n}}\Gamma(\alpha + s + n) = \sum \limits_{n=1}^{\infty} \frac{\Gamma(\alpha + 1) \Gamma(\alpha + s + n)}{\Gamma(\alpha + n + 1)2^{\alpha+s+n}}.
\end{equation}
Introducing this result into Eq.~\eqref{eq:i_nu_s1_s2},
\begin{equation}
\begin{aligned}
I(\nu, s_1, s_2) &= \frac{1}{\nu+1} \left( I'(\nu+s_2,s_1) + I'(\nu+s_1,s_2)  \right) \\
&= \frac{1}{\nu+1} \sum \limits_{n=1}^{\infty} \left( \frac{\Gamma(\nu+s_2 + 1) \Gamma(\nu+s_1 + s_2 + n)}{\Gamma(\nu+s_2 + n + 1)2^{\nu+s_1+s_2+n}} + \frac{\Gamma(\nu+s_1 + 1) \Gamma(\nu+s_1 + s_2 + n)}{\Gamma(\nu+s_1 + n + 1)2^{\nu+s_1+s_2+n}} \right) \\
&= \frac{1}{(\nu+1)2^{\nu+s_1+s_2}} \sum \limits_{n=1}^{\infty} \frac{\Gamma(\nu+s_1 + s_2 + n)}{2^n} \left( \frac{\Gamma(\nu+s_2 + 1)}{\Gamma(\nu+s_2 + n + 1)} + \frac{\Gamma(\nu+s_1 + 1) }{\Gamma(\nu+s_1 + n + 1)} \right).
\end{aligned}
\end{equation}
Finally, Eq.~\eqref{eq:av_kappap_b2} becomes
\begin{equation}\label{eq:b_2_exact}
\begin{aligned}
\langle \kappa' \rangle &=  \kappa_0 \frac{2 (\gamma - 1)^2}{\beta^2 \Gamma \left(1- \frac{1}{\beta}\right)} \int \limits_0^\infty \omega^{2 \eta - 3 - 1/\beta} \Gamma(1-\eta, \omega)\Gamma(2-\eta, \omega) \text{d}\omega \\
&= \kappa_0 \frac{2 (\gamma - 1)^2}{\beta^2 \Gamma \left(1- \frac{1}{\beta}\right)} I(2 \eta - 3 - 1/\beta, 1- \eta, 2-\eta) \\
&= \frac{2^{1+\frac{1}{\beta}} (\gamma - 1)^2  \kappa_0}{\beta \Gamma \left(1- \frac{1}{\beta}\right)(2\gamma-3)} \sum \limits_{n=1}^{\infty} \frac{\Gamma \left( n - \frac{1}{\beta} \right)}{2^n} \left( \frac{\Gamma \left(\frac{\gamma - 2}{\beta} + 1\right)}{\Gamma \left(\frac{\gamma - 2}{\beta}+ n + 1 \right)} + \frac{\Gamma \left(\frac{\gamma-2}{\beta} \right) }{\Gamma \left(\frac{\gamma-2}{\beta} + n \right)} \right).
\end{aligned}
\end{equation}

\begin{figure}[h!]
\centering
\includegraphics[width=0.4\columnwidth]{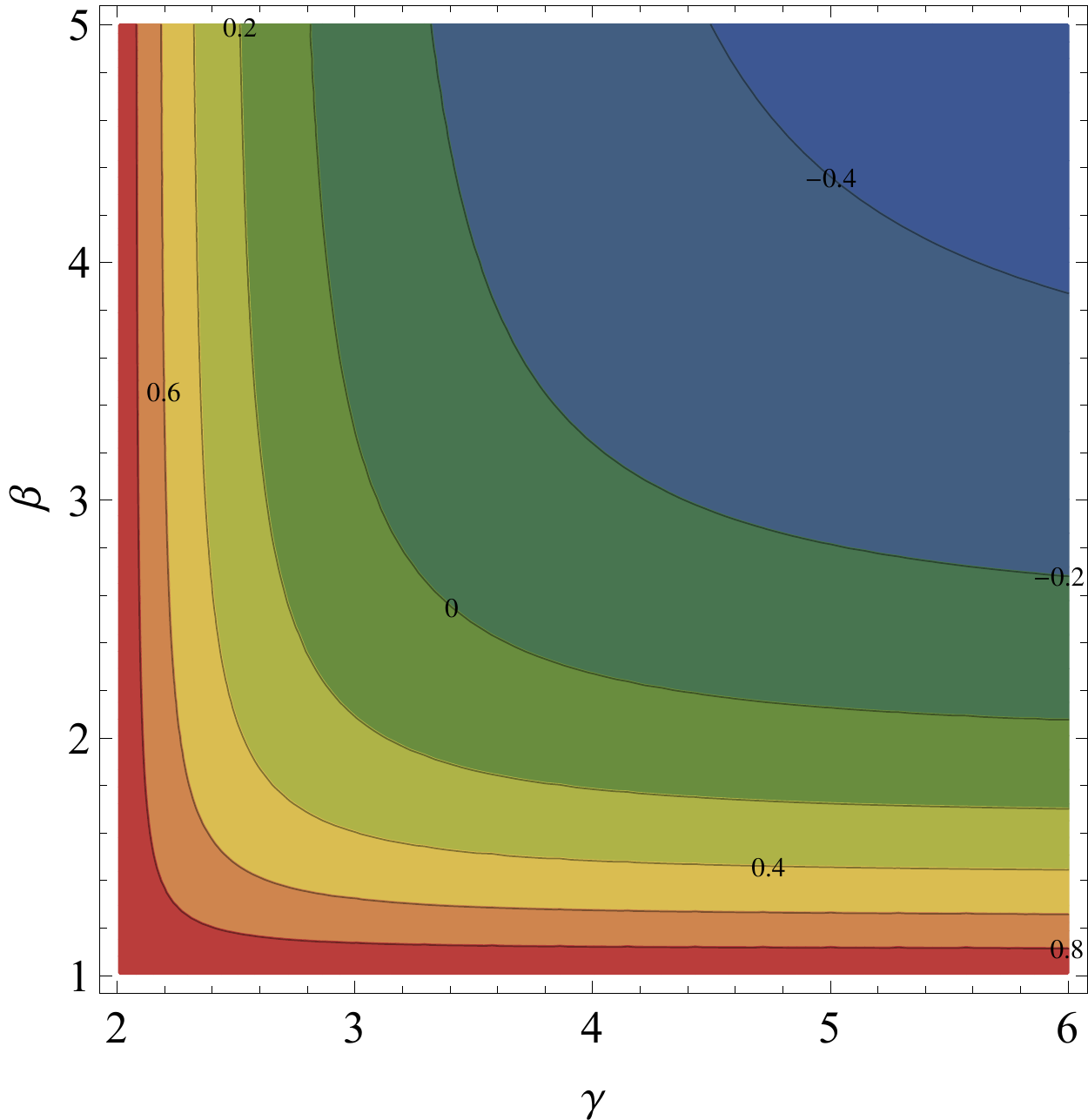}
\caption{\label{fig:nu_exact} \textbf{Connectivity phase diagram.} Exact value of $\nu$ as a function of $\beta$ and $\gamma$ according to Eq.~\eqref{eq:nu_exact}. The exact solution agrees with the solution in the power-law approximation (see next subsection) for large values of $\beta$ or $\gamma$.}
\end{figure}

Using this result, Eq.~\eqref{eq:av_k_av_kappa} and $\kappa_0 = \langle \kappa \rangle (\gamma - 2)/(\gamma - 1)$ we can write an expression for the exponent $\nu$ (defined by the expression $\langle k' \rangle = r^{\nu} \langle k \rangle$):
\begin{equation}\label{eq:nu_exact}
\nu = \frac{2}{\ln 2} \ln \left[ \frac{2^{1+\frac{1}{\beta}} (\gamma - 1)(\gamma - 2)}{\beta \Gamma \left(1- \frac{1}{\beta}\right)(2\gamma-3)} \sum \limits_{n=1}^{\infty} \frac{\Gamma \left( n - \frac{1}{\beta} \right)}{2^n} \left( \frac{\Gamma \left(\frac{\gamma - 2}{\beta} + 1\right)}{\Gamma \left(\frac{\gamma - 2}{\beta}+ n + 1 \right)} + \frac{\Gamma \left(\frac{\gamma-2}{\beta} \right) }{\Gamma \left(\frac{\gamma-2}{\beta} + n \right)} \right) \right] - 1.
\end{equation}
The above result is shown in Fig.~\ref{fig:nu_exact}.

\subsubsection*{Solution in the power-law approximation}
From Eq.~\eqref{eq:b_2_exact}, we see that the exact solution for large $r$ can be extremely convoluted, thus making the limit $r \to \infty$ inaccessible. However, if we consider that $\rho_{\kappa'} (\kappa')$ is a power-law (which is a reasonable approximation if $\eta < 3$, as discussed above), the computation of $\langle \kappa' \rangle$ becomes simpler. Under this assumption, $z'$ are also power-law distributed with exponent $-\eta$, that is,

\begin{equation}\label{eq:assumption_zp}
\rho_{z'}(z') = C' z'^{- \eta}, \quad C' = \frac{\gamma - 1 }{\beta \kappa_0'^{1 - \gamma}}.
\end{equation}
We study two cases separately:

\begin{itemize}
\item[i.] $1 < \eta < 2$: In this case, we determine the value of $C'$ and, with it, $\langle \kappa' \rangle = \kappa'_0 \frac{\gamma-1}{\gamma-2}$. If the assumption in Eq.~\eqref{eq:assumption_zp} is correct, $\hat{\rho}_{z'}(s)$ must behave as~\citep{handelsman}
\begin{equation}
\hat{\rho}_{z'}(s) = 1 + C' s^{\eta-1} \Gamma(1-\eta), \quad s \to 0^+.
\end{equation}
According to Eqs.~\eqref{eq:rho_hat_zp} and \eqref{eq:rho_hat_z_c},
\begin{equation}
\begin{aligned}\label{eq:rho_hat_zp_expansion}
\hat{\rho}_{z'}(s) &=\left[ C s^{\eta-1} \Gamma \left( 1 - \eta, s z_0 \right) \right]^r = \left[ C s^{\eta-1} \Gamma \left( 1 - \eta \right) \left( 1 - (sz_0)^{1-\eta} e^{-sz_0} \sum \limits_{n=0}^{\infty} \frac{(sz_0)^n}{\Gamma ( 2 - \eta + n)} \right) \right]^r \\
&= \left[ C \Gamma \left( 1 - \eta \right) \left( s^{\eta-1} - z_0^{1-\eta} e^{-sz_0} \sum \limits_{n=0}^{\infty} \frac{(sz_0)^n}{\Gamma ( 2 - \eta + n)} \right) \right]^r \\
 &\to \left[ C \Gamma \left( 1 - \eta \right) \left( s^{\eta-1} - z_0^{1-\eta} \sum \limits_{n=0}^{\infty} \frac{(sz_0)^n}{\Gamma ( 2 - \eta + n)} \right) \right]^r, \quad s \to 0^+.
\end{aligned}
\end{equation}
In the above expression, we see that the term that does not depend on $s$ is given by the product of the $r$ terms with $n=0$, whereas the term of order $s^{\eta-1}$ is given by the sum of the $r$ products of $s^{\eta-1}$ with the remaining $r-1$ terms with $n=0$. Thus, we find
\begin{equation}
\begin{aligned}
\hat{\rho}_{z'}(s) &= \left[ C \Gamma \left( 1 - \eta \right) \left( s^{\eta-1} - z_0^{1-\eta} \sum \limits_{n=0}^{\infty} \frac{(sz_0)^n}{\Gamma ( 2 - \eta + n)} \right) \right]^r \\
&\to C^r \Gamma^r \left( 1 - \eta \right) \left[(-1)^r \frac{z_0^{r(1-\eta)}}{\Gamma^r ( 2 - \eta)} + r (-1)^{r-1} s^{\eta-1} \frac{z_0^{(r-1)(1-\eta)}}{\Gamma^{r-1} ( 2 - \eta)} \right] \\
&= \left( \frac{\gamma-1}{\beta \kappa_{0}^{1-\gamma}} \right)^r \Gamma^r \left( 1 - \eta \right) \left[(-1)^r \frac{z_0^{r(1-\eta)}}{( 1 - \eta)^r \Gamma^r ( 1 - \eta)} + r (-1)^{r-1} s^{\eta-1} \frac{z_0^{(r-1)(1-\eta)}}{( 1 - \eta)^{r-1} \Gamma^{r-1} ( 1 - \eta)} \right] \\
&= \left( \frac{\eta-1}{z_{0}^{1-\eta}} \right)^r \left[ \frac{z_0^{r(1-\eta)}}{(\eta-1)^r} + r s^{\eta-1} \Gamma \left( 1 - \eta \right) \frac{z_0^{(r-1)(1-\eta)}}{(\eta-1)^{r-1}} \right] \\
&= 1 + r \frac{\eta-1}{z_{0}^{1-\eta}} s^{\eta-1} \Gamma \left( 1 - \eta \right).
\end{aligned}
\end{equation}
We can now identify $C'$ as
\begin{equation}
C' = \frac{\gamma - 1 }{\beta \kappa_0'^{1 - \gamma}} = r \frac{\eta-1}{z_{0}^{1-\eta}} = r \frac{\gamma-1}{\beta \kappa_{0}^{1-\gamma}},
\end{equation}
so
\begin{equation}
\kappa_0' = r^{\frac{1}{\gamma-1}} \kappa_0
\end{equation}
and
\begin{equation}
\langle \kappa' \rangle = r^{\frac{1}{\gamma-1}} \langle \kappa \rangle.
\end{equation}
Finally, plugging this result into Eq.~\eqref{eq:av_k_av_kappa},
\begin{equation}\label{eq:approx_avk_upper}
\langle k' \rangle = C_0 \frac{\mu}{r} r^{\frac{2}{\gamma-1}} \langle \kappa \rangle^2 = r^{\frac{2}{\gamma-1} - 1} \langle k \rangle \to 
\left\lbrace 
\begin{array}{c l}
\infty & \gamma < 3 \\
cte. & \gamma = 3 \\
0 & \gamma > 3
\end{array}
\right.
\end{equation}
\item[ii.] $2 < \eta < 3$: This case is much simpler, since $\langle z \rangle$ and hence $\langle z' \rangle$ are finite and can be easily computed. Indeed, given that $\langle z' \rangle = r \langle z \rangle$, we see that
\begin{equation}
\begin{aligned}
\langle \kappa' \rangle &= \frac{\gamma - 1}{\gamma - 2} \kappa'_0 = \frac{\gamma - 1}{\gamma - 2} \left(z_0'\right)^{1/\beta} = \frac{\gamma - 1}{\gamma - 2} \left( \frac{\eta - 2}{\eta - 1} \langle z' \rangle \right)^{1/\beta} = \frac{\gamma - 1}{\gamma - 2} \left( \frac{\eta - 2}{\eta - 1} r \langle z \rangle \right)^{1/\beta} \\
&= \frac{\gamma - 1}{\gamma - 2} \left( r\frac{\eta - 2}{\eta - 1} \frac{\eta - 1}{\eta - 2} z_0 \right)^{1/\beta} = \frac{\gamma - 1}{\gamma - 2} \left( r \kappa_0^\beta \right)^{1/\beta} = r^{1/\beta} \langle \kappa \rangle.
\end{aligned}
\end{equation}
This result and Eq.~\eqref{eq:av_k_av_kappa} together imply
\begin{equation}\label{eq:approx_avk_lower}
\langle k' \rangle = C_0 \frac{\mu}{r} r^{2/\beta} \langle \kappa \rangle^2 = r^{2/\beta - 1} \langle k \rangle \to 
\left\lbrace 
\begin{array}{c l}
\infty & \beta < 2 \\
cte. & \beta = 2 \\
0 & \beta > 2
\end{array}
\right.
\end{equation}

\end{itemize}

Both solutions, Eqs.~\eqref{eq:approx_avk_upper} and \eqref{eq:approx_avk_lower}, are equivalent at $\eta=2$, since
\begin{equation}
\eta = 2 \Rightarrow \frac{\gamma-1}{\beta} = 1 \Rightarrow \beta = \gamma-1.
\end{equation}
Therefore, we can conclude that the network flows towards a fully connected graph if $\gamma<3$ or $\beta<2$. The line $\gamma = 3$ and $\beta > 2$ or $\beta = 2$ and $\gamma > 3$ is an unstable fixed point, whereas $\langle k \rangle \to 0$ if $\gamma > 3$ and $\beta > 2$. Notice that this assertion is only valid under the assumption in Eq.~\eqref{eq:assumption_zp}, which is not true in general. However, we expect it to be a good approximation of the flow's behaviour as $r \to \infty$.

\subsection{Mapping to hyperbolic space and the partition function}

In this section, we show how the RGN presented in this work can be described in the formalism of statistical physics. As explained in Appendix~\ref{app:met}, using the mapping to hyperbolic space, the connection probability, Eq.~\eqref{eq:pij_original}, becomes
\begin{equation}\label{eq:pij_h2}
p_{mn} = \frac{1}{1 + e^{\frac{\beta}{2} (x_{mn} - R_{\mathbb{H}^2})}},
\end{equation}
where $x_{mn} = r_{m} + r_{n} + 2 \ln \frac{\Delta \theta_{mn}}{2}$ is a good approximation to the hyperbolic distance between two points with coordinates $(r_{m}, \theta_{m})$ and $(r_{n}, \theta_{n})$ in the native representation of hyperbolic space.

Now, let $a_{mn} = 1$ if the link between nodes $m$ and $n$ exists and $a_{mn} = 0$ otherwise; Eq.~\eqref{eq:pij_h2} can be written as
\begin{equation}
p_{mn} \equiv P(a_{mn}) = \frac{e^{- \beta \frac{a_{mn}}{2} (x_{mn} - R_{\mathbb{H}^2})}}{1 + e^{-\frac{\beta}{2} (x_{mn} - R_{\mathbb{H}^2})}},
\end{equation}
which means that, in the $\mathbb{H}^2$ model, every pair of nodes represents a fermionic state of energy $x_{mn}/2$ in the grand-canonical ensemble with $R_{\mathbb{H}^2}/2$ playing the role of the chemical potential. Indeed, since a network can be represented by the set $\lbrace a_{mn} \rbrace$, the likelihood of a given network is given by
\begin{equation}
P(\lbrace a_{mn} \rbrace) = \prod \limits_{m < n} \frac{e^{- \beta \frac{a_{mn}}{2} (x_{mn} - R_{\mathbb{H}^2})}}{1 + e^{-\frac{\beta}{2} (x_{mn} - R_{\mathbb{H}^2})}},
\end{equation}
that is, by the probability of the corresponding microstate of the gas of non-interacting fermions. The partition function of the system is
\begin{equation}
Z = \prod \limits_{m < n} \sum \limits_{a_{mn} = 0}^{1} e^{-\beta \frac{a_{mn}}{2} (x_{mn} - R_{\mathbb{H}^2})} = \prod \limits_{m < n} \left( 1 + e^{-\frac{\beta}{2} (x_{mn} - R_{\mathbb{H}^2})} \right).
\end{equation}
When we apply the renormalization transformation, every node $m$ ($n$) is mapped to a supernode $i$ ($j$). We can rearrange the terms in the partition function according to such mapping as
\begin{equation}\label{eq:partition_separated}
Z = \prod \limits_{i = 1}^{\left \lfloor \frac{N}{r} \right \rfloor} \prod \limits_{t = 1}^{\frac{r(r-1)}{2}} \left( 1 + e^{-\frac{\beta}{2} (x_{t} - R_{\mathbb{H}^2})} \right) \prod \limits_{i < j} \prod \limits_{e = 1}^{r^2} \left( 1 + e^{-\frac{\beta}{2} (x_{e} - R_{\mathbb{H}^2})} \right).
\end{equation}
The first double product in the above expression corresponds to the partial sum over the links among the nodes within every supernode $i$ (hence, there are $N(r-1)/2$ such terms), whereas the second double product represents the partial sum over the links among nodes in different supernodes $i$ and $j$; thus, it contains $(N^2-Nr)/2$ terms. According to Eqs.~\eqref{eq:mapping_Rh2} and \eqref{eq:mapping_r},
\begin{equation}
e^{-\frac{\beta}{2} (x_{mn} - R_{\mathbb{H}^2})} = \left( \frac{\mu \kappa_{m} \kappa_{n}}{R \Delta \theta_{mn}} \right)^{\beta},
\end{equation}
so the rightmost term in Eq.~\eqref{eq:partition_separated} reads
\begin{equation}
\prod \limits_{i < j} \prod \limits_{e = 1}^{r^2} \left( 1 + e^{-\frac{\beta}{2} (x_{e} - R_{\mathbb{H}^2})} \right) = \prod \limits_{i < j} \prod \limits_{e = 1}^{r^2} \left( 1 + \left( \frac{\mu (\kappa_{m} \kappa_{n})_{e}}{R \Delta \theta_{e}} \right)^{\beta} \right) = \prod \limits_{i < j} \left( 1 + \Phi_{ij}' \right),
\end{equation}
where $\Phi_{ij}'$ is given by Eq.~\eqref{eq:phi_exact}. Using Eq.~\eqref{eq:phi_approx} and Eq.~\eqref{eq:params_transf}, which is fulfilled with the RG transformations Eqs.~\eqref{eq:kappa_p} and \eqref{eq:theta_p}, yields
\begin{equation}
\begin{aligned}
\prod \limits_{i < j} \prod \limits_{e = 1}^{r^2} \left( 1 + e^{-\frac{\beta}{2} (x_{e} - R_{\mathbb{H}^2})} \right) &\approx \prod \limits_{i < j} \left( 1 + \left( \frac{\mu}{R \Delta \theta} \right)^{\beta} \sum \limits_{e=1}^{r^2} \left( \kappa_{m} \kappa_{n} \right)_{e}^{\beta} \right) \\
&= \prod \limits_{i < j} \left( 1 + \left( \frac{\mu' \kappa'_{i} \kappa'_{j}}{R' \Delta \theta'_{ij}} \right)^{\beta} \right) = \prod \limits_{i < j} \left( 1 + e^{-\frac{\beta}{2} (x'_{ij} - R'_{\mathbb{H}^2})} \right) = Z'.
\end{aligned}
\end{equation}
The leftmost term in Eq.~\eqref{eq:partition_separated} can be written as
\begin{equation}
\prod \limits_{i = 1}^{\left \lfloor \frac{N}{r} \right \rfloor} \prod \limits_{t = 1}^{\frac{r(r-1)}{2}} \left( 1 + e^{-\frac{\beta}{2} (x_{t} - R_{\mathbb{H}^2})} \right) = \prod \limits_{i = 1}^{\left \lfloor \frac{N}{r} \right \rfloor} \prod \limits_{t = 1}^{\frac{r(r-1)}{2}} \left( 1 +  \left( \frac{\mu (\kappa_{m} \kappa_{n})_{t}}{R \Delta \theta_{t}} \right)^{\beta} \right).
\end{equation}
In the particular case of $r=2$, we integrate consecutive nodes separated by a typical angular distance $\Delta \theta_t \approx 2 \pi/N$. Hence, $R \Delta \theta_t \approx 1$, so
\begin{equation}
\prod \limits_{i = 1}^{\left \lfloor \frac{N}{r} \right \rfloor} \prod \limits_{t = 1}^{\frac{r(r-1)}{2}} \left( 1 + e^{-\frac{\beta}{2} (x_{t} - R_{\mathbb{H}^2})} \right) \approx e^{\sum\limits_{i=1}^{\lfloor N/2 \rfloor} \ln \left( 1 + (\mu (\kappa_m \kappa_n)_{i})^{\beta} \right)} \approx e^{\frac{N}{2} \left \langle \ln \left( 1 + (\mu \kappa_m \kappa_n)^{\beta} \right) \right \rangle}.
\end{equation}
Defining
\begin{equation}
\zeta \equiv e^{\left \langle \ln \left( 1 + (\mu \kappa_m \kappa_n)^{\beta} \right) \right \rangle} = e^{\int \ln \left( 1 + (\mu \kappa_m \kappa_n)^{\beta} \right) \rho(\kappa_m) \rho(\kappa_n) d \kappa_m d \kappa_n}
\end{equation}
we can write Eq.~\eqref{eq:partition_separated} as
\begin{equation}
Z = \zeta^{N/2} Z',
\end{equation}
where $Z' = \sum_{\lbrace a_{ij} \rbrace} e^{- \beta H' (\lbrace a_{ij}\rbrace )}$.

\subsection{Local vs. global properties}
\begin{figure}[h!]
\centering
\includegraphics[width=1.0\columnwidth]{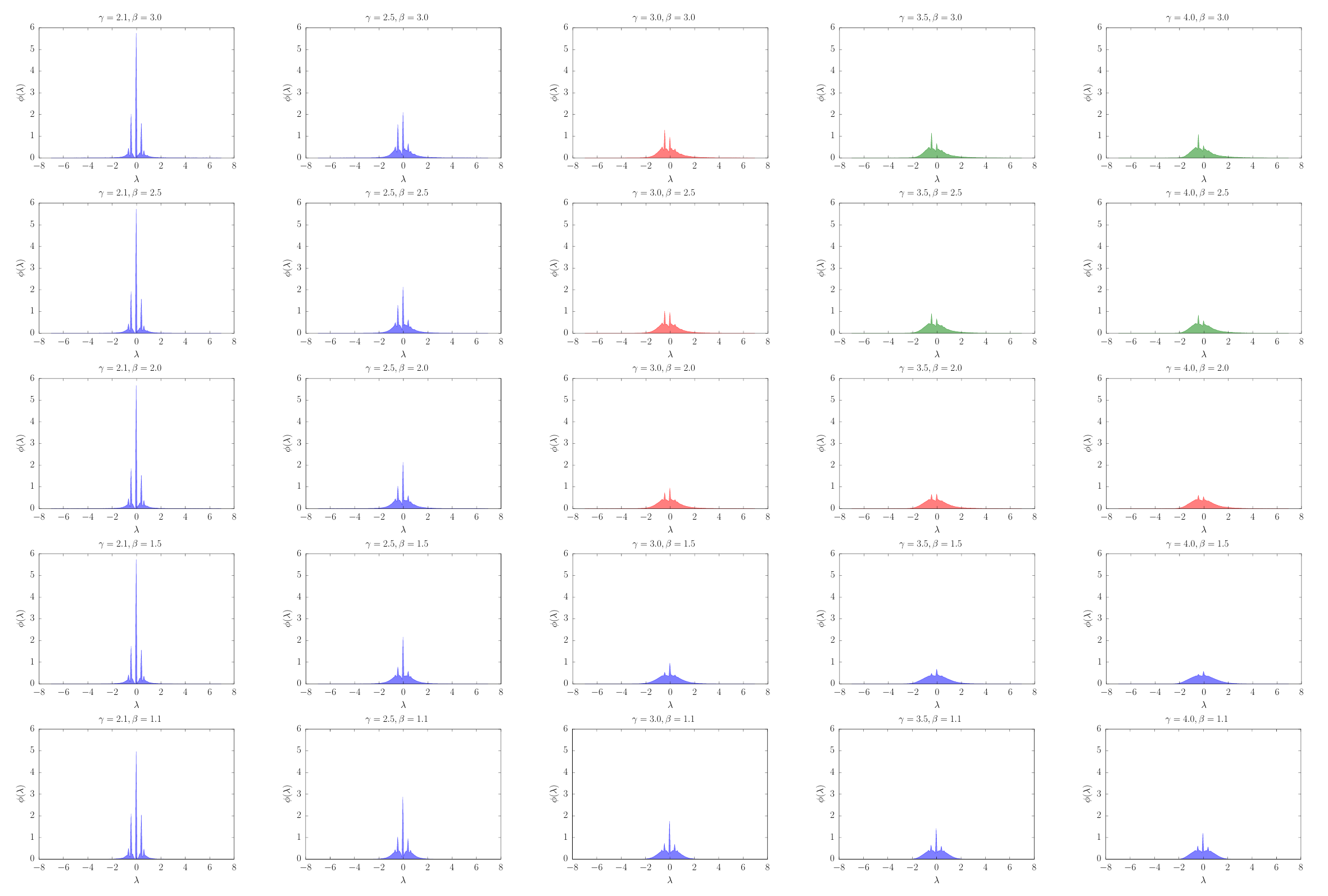}
\caption{\label{fig:adjacency} \textbf{Eigenvalues of adjacency matrices.} Every plot represents a histogram of the eigenvalues (divided by $\sqrt{\langle \kappa \rangle}$) of the adjacency matrices of 100 synthetic networks of size $N = 1000$ and $\langle \kappa \rangle = 5$ for a particular set of values $(\gamma, \beta)$. The order of the plots corresponds to that of the phase diagram Fig.~\ref{fig3}{\bf B}. Notice how the RG analysis correctly predicts the dependence of the spectra on $\gamma$ only on the top-left corner of the figure, as well as the independence on $\gamma$ on the bottom-right region.}
\end{figure}

\begin{figure}[h!]
\centering
\includegraphics[width=1.0\columnwidth]{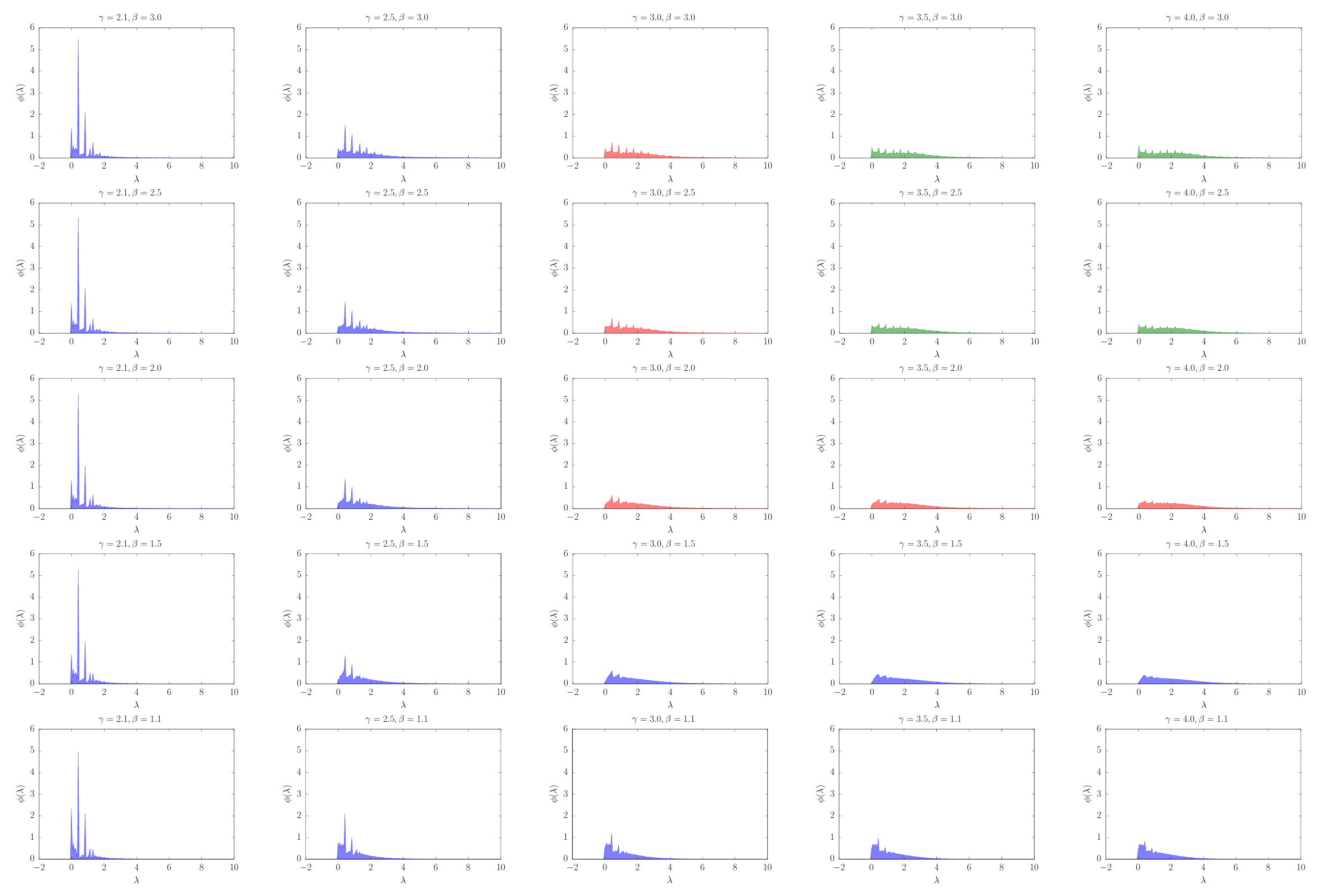}
\caption{\label{fig:laplacian} \textbf{Eigenvalues of laplacian matrices.}Every plot represents a histogram of the eigenvalues (divided by $\sqrt{\langle \kappa \rangle}$) of the laplacian matrices of 100 synthetic networks of size $N = 1000$ and $\langle \kappa \rangle = 5$ for a particular set of values $(\gamma, \beta)$. The order of the plots corresponds to that of the phase diagram Fig.~\ref{fig3}{\bf B}. Notice how the RG analysis correctly predicts the dependence of the spectra on $\gamma$ only on the top-left corner of the figure, as well as the independence on $\gamma$ on the bottom-right region.}
\end{figure}

\begin{figure}[h!]
\centering
\includegraphics[width=0.49\columnwidth]{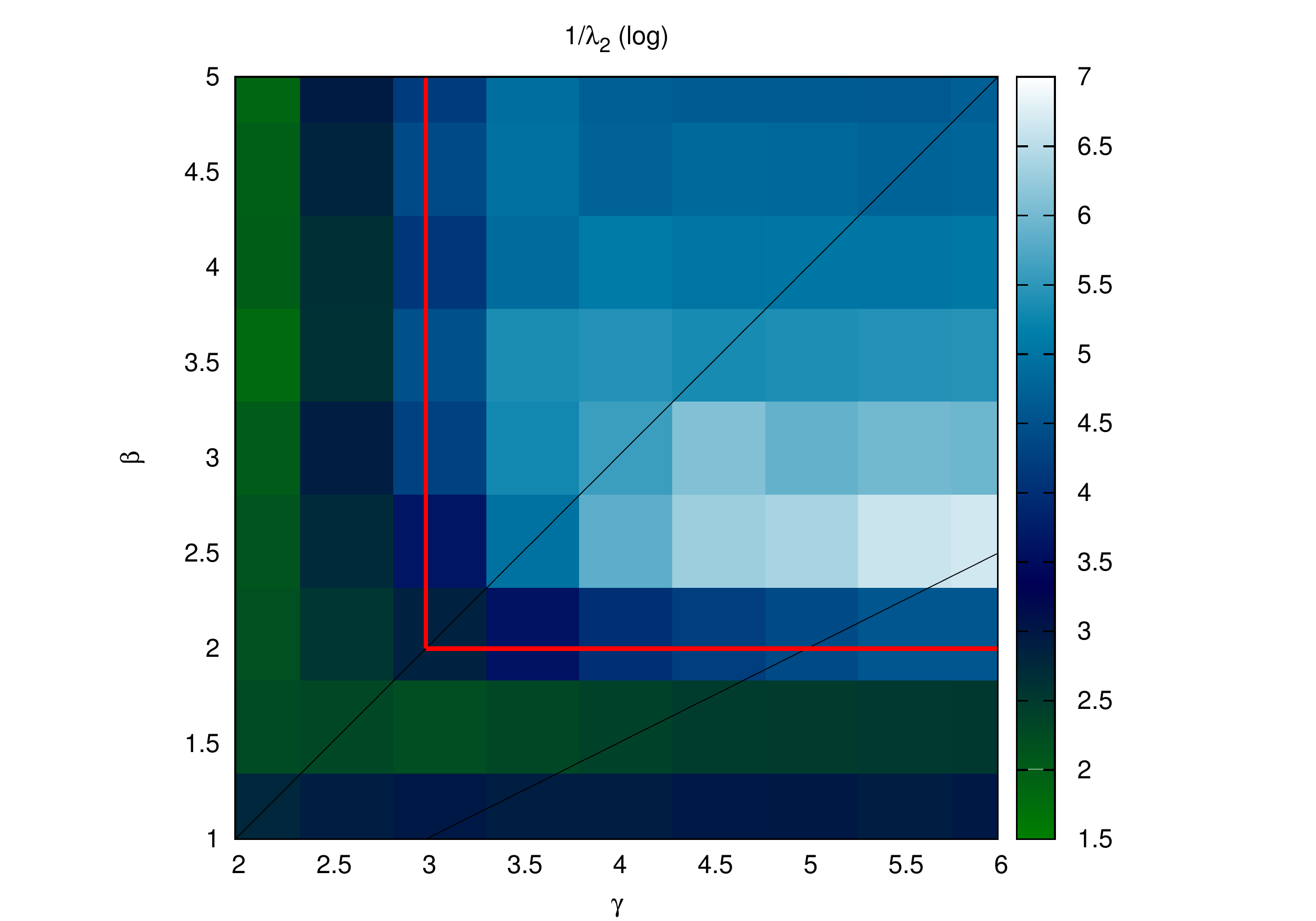}
\includegraphics[width=0.49\columnwidth]{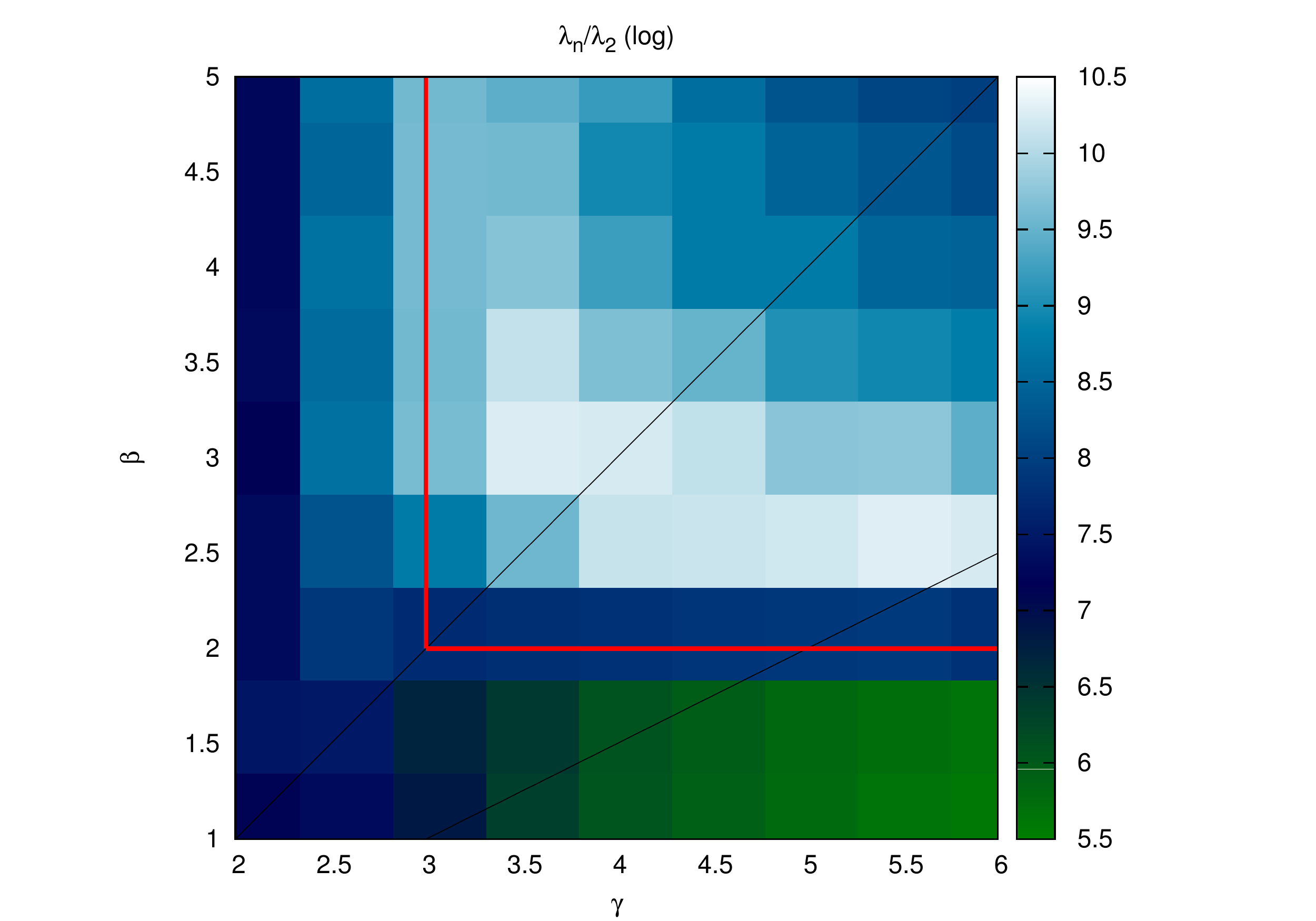}
\caption{\label{fig:algebraic_connectivity} \textbf{Diffusion time and synchronization stability.} \textbf{Left:} Logarithm of the diffusion time (inverse of the algebraic connectivity or first non-null eigenvalue of the laplacian $\lambda_2$) of networks of size $N = 1000$ and $\langle \kappa \rangle = 5$ averaged over 100 realizations. \textbf{Bottom:} Logarithm of the quotient $\lambda_n/\lambda_2$ (this quantity is related to the stability of synchronization processes on networks; it gives the time that the system needs to get back to the stable synchronized state after a perturbation occurred). In both plots, we can see the similarities with Fig.~\ref{fig3}{\bf B}.}
\end{figure}

In the $\mathbb{S}^1$ model, we impose three parameters, $\gamma, \beta$ and $\langle \kappa \rangle$, all three related to local properties of nodes (degree and clustering). However, the RG flow of observables like the average degree should be related to global properties of the system; indeed, we would expect two networks with similar average degree flows to exhibit similarities at the global scale as well, whereas two networks with very different RG trajectories (even in the same phase, i.e., flowing towards the same fixed point) should be easier to distinguish by looking at their global properties. To check this hypothesis, we have generated synthetic networks with different values of $\gamma$ and $\beta$ and compared the eigenvalues of both the adjacency and laplacian matrices. The results are shown in Figs.~\ref{fig:adjacency},~\ref{fig:laplacian} and~\ref{fig:algebraic_connectivity}. As we see, the RG analysis of the model allows us to assess the stability of the global properties of networks against perturbations of their local ones, and hence the importance of clustering and degree heterogeneity on a given system.
\newpage

\clearpage

\section{Mini-me network replicas}\label{app:dyn}
\begin{figure}[h!]
\centering
\includegraphics[width=0.9\columnwidth]{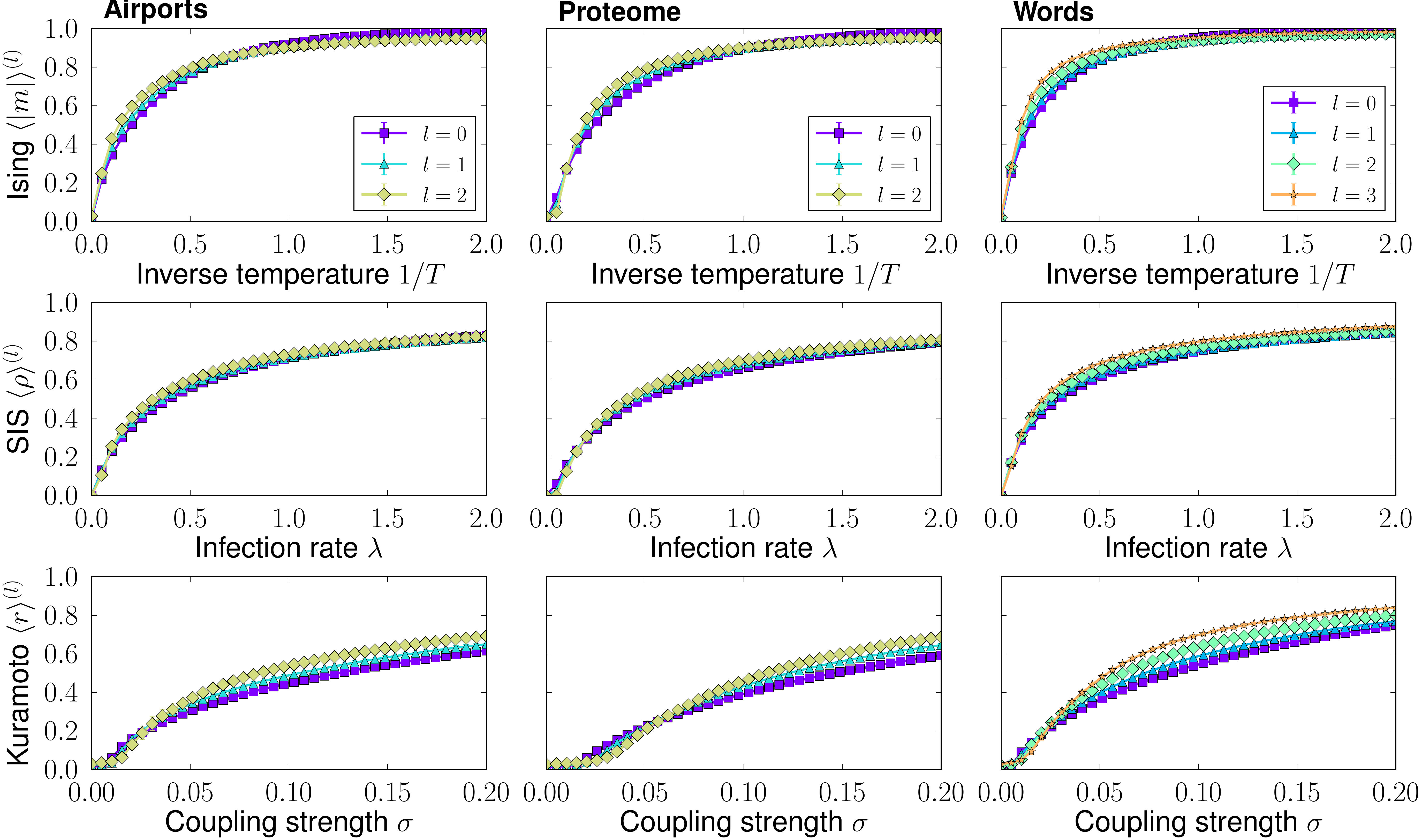}
\caption{\label{fig:dynamics}{\bf Dynamics on the Mini-me replicas.} Each column shows the order parameters versus the control parameters of different dynamical processes on the original and Mini-me replicas of the Airports network (left), the Proteome network (middle) and the Words network (right) with $r = 2$, that is, every value of $l$ identifies a network $2^{l}$ times smaller than the original one. All points show the results averaged over 100 simulations. Error bars indicate the fluctuations of the order parameters. {\bf Top:} Magnetization $\langle | m | \rangle^{(l)}$ of the Ising model as a function of the inverse temperature $1/T$. {\bf Middle:} Prevalence $\langle \rho \rangle^{(l)}$ of the SIS model as a function of the infection rate $\lambda$. {\bf Bottom:} Coherence $\langle r \rangle^{(l)}$ of the Kuramoto model as a function of the coupling strength $\sigma$. In all cases, the curves of the smaller-scale replicas are extremely similar to the results obtained on the original networks.}
\end{figure}
\clearpage 

\section{Multiscale navigation networks}\label{app:mul}
\begin{figure}[h!]
\centering
\includegraphics[width=1.0\columnwidth]{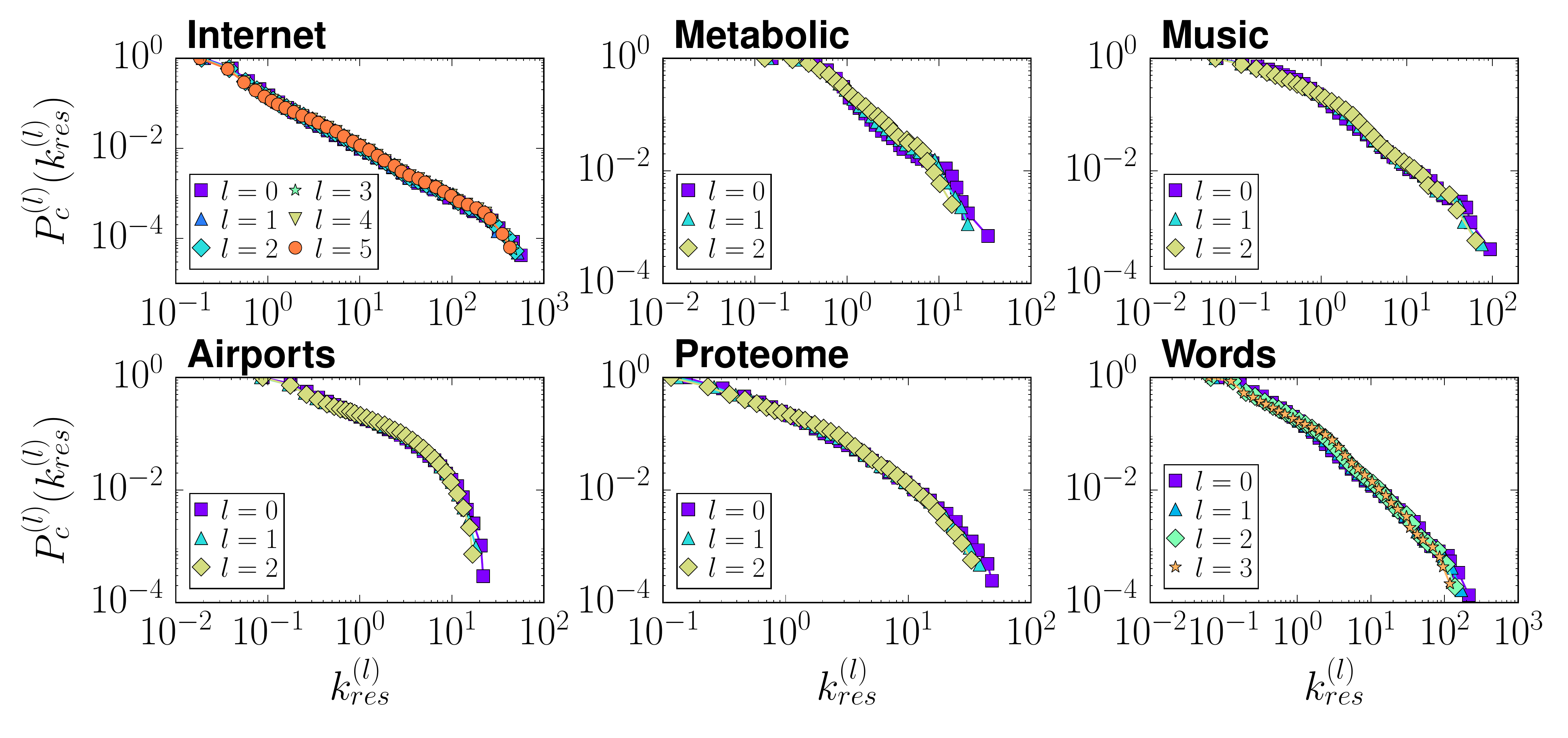}
\caption{\label{fig:nav_pk} \textbf{Complementary cumulative degree distributions.} Every curve represents the complementary cumulative degree distribution of a given layer in the multiscale navigation shell.}
\end{figure}

\begin{figure}[h!]
\centering
\includegraphics[width=1.0\columnwidth]{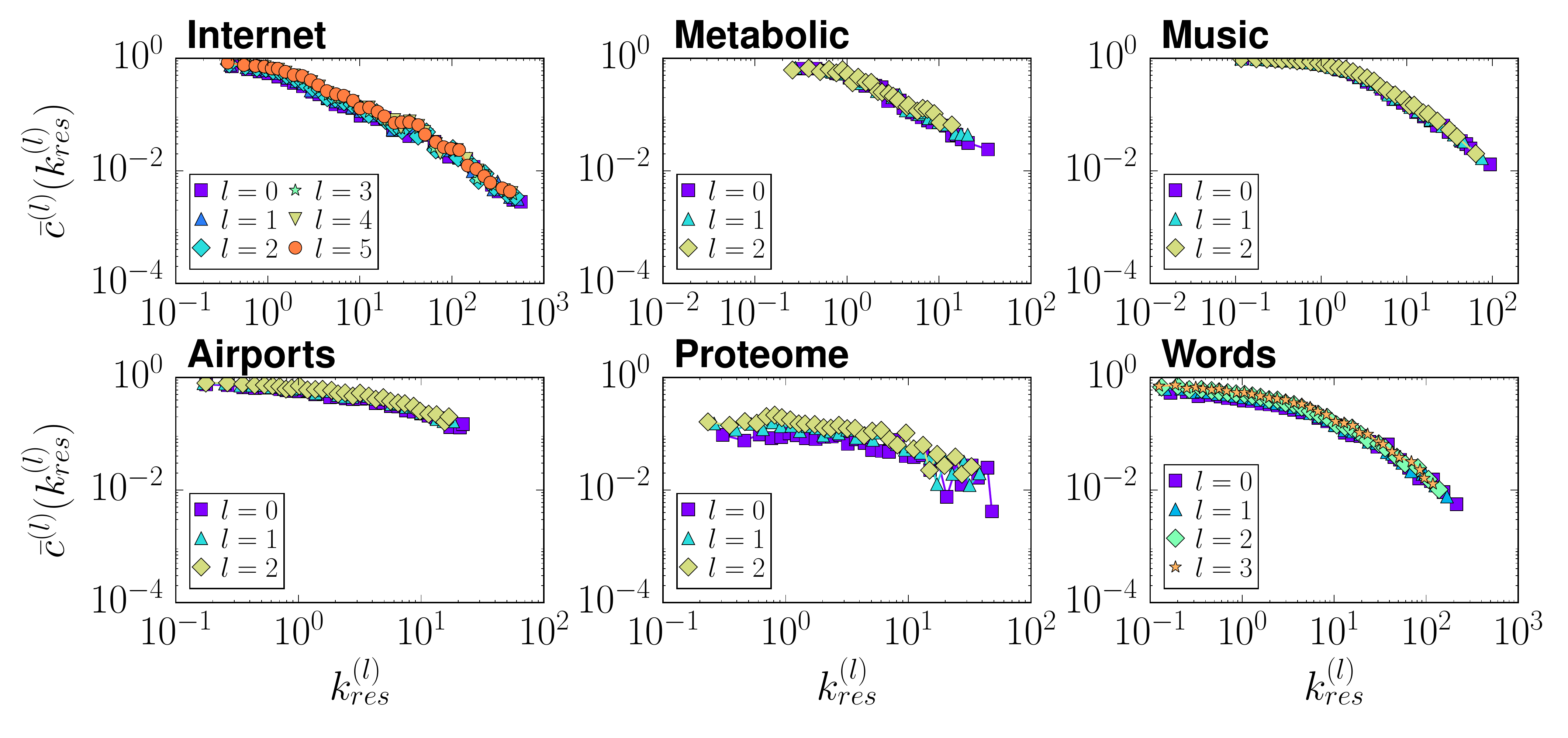}
\caption{\label{fig:nav_ck} \textbf{Clustering spectra.} Every curve represents the clustering spectrum of a given layer in the multiscale navigation shell.}
\end{figure}

This section includes some results showing the topological properties of the coarse-grained for navigation networks; Fig.~\ref{fig:nav_pk} shows the complementary cumulative degree distributions, whereas Fig.~\ref{fig:nav_ck} contains their clustering spectra.

\begin{figure}[h!]
\centering
\includegraphics[width=1.0\columnwidth]{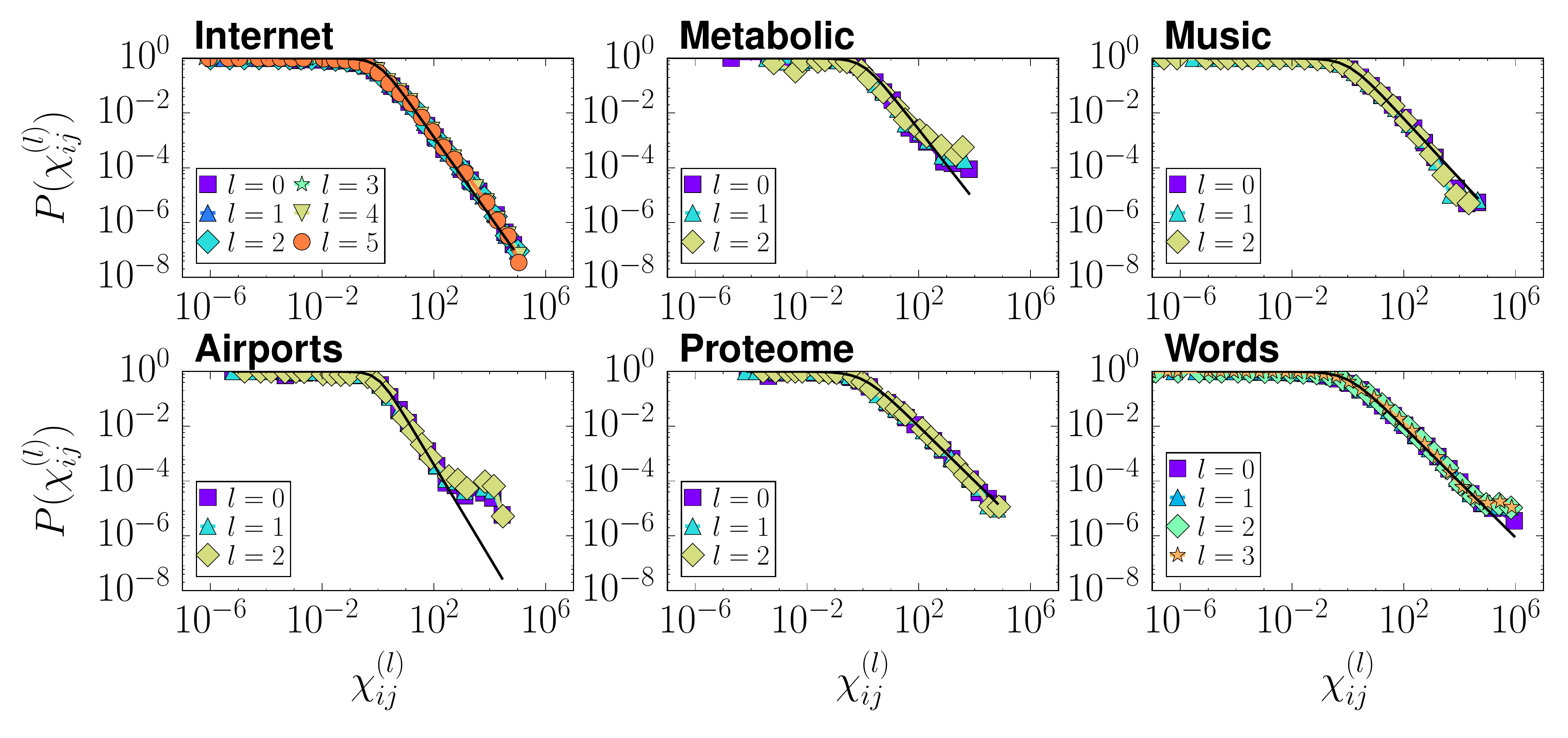}
\caption{\label{fig:conn_prob_navigation}  \textbf{Empirical connection probabilities.} Fraction of connected pairs within a given range of $\chi_{ij}^{(l)}$ for the six real-world networks and their coarse-grained for navigation versions. The black curve is the theoretic connection probability.}
\end{figure}

We also present the empirical connection probabilities of the networks after the coarse-graining for navigation (in which pairs of nodes are merged together into a supernode only if they are connected) in Fig.~\ref{fig:conn_prob_navigation}. Notice that the congruency with the underlying metric space is preserved even is the sizes of the blocks are different.

\end{document}